\documentclass[floatfix,pra,twocolumn,showpacs,amsmath,amssymb]{revtex4-1}
\usepackage{times}
\usepackage{latexsym}
\usepackage{graphicx}
\usepackage[english]{babel}
\usepackage{verbatim,times,bbm}
\usepackage{color}
\usepackage{hyperref}

\begin{document}

\title{Enhancing Quantum Effects via Periodic Modulations in Optomechanical Systems}
\author{Alessandro Farace and Vittorio Giovannetti}
\affiliation{NEST-CNR-INFM \& Scuola Normale Superiore, Piazza dei Cavalieri 7, I-56126 Pisa, Italy}

\begin{abstract}

Parametrically modulated optomechanical systems have been recently proposed as a simple and efficient setting for the quantum control of a micromechanical oscillator: relevant possibilities include the generation of squeezing in the oscillator position (or momentum) and the enhancement of entanglement between mechanical and radiation modes. In this paper we further investigate this new modulation regime, considering an optomechanical system with one or more parameters being modulated over time. We first apply a sinusoidal modulation of the mechanical frequency and characterize the optimal regime in which the visibility of purely quantum effects is maximal. We then introduce a second modulation on the input laser intensity and analyze the interplay between the two. We find that an interference pattern shows up, so that different choices of the relative phase between the two modulations can either enhance or cancel the desired quantum effects, opening new possibilities for optimal quantum control strategies.

\end{abstract}


\maketitle

\section{Introduction}
\label{SecIntro}
Theoretical studies and huge technological progresses over the last decades made it possible to reach a considerable level of control over quantum states of matter in a large variety of physical systems, ranging from photons, electrons and atoms to bigger solid state systems such as quantum dots and superconducting circuits. This opened the possibility for novel tests of quantum mechanics and allowed, among other things, to take important steps forward in investigating the quantum regime of macroscopic objects. In this perspective, one of the main goals in today quantum science is controlling nano- and micromechanical oscillators at the quantum level.

Quantum optomechanics \cite{Marquardt1, Aspelmeyer1, Kippenberg1, Milburn1}, i.e. studying and engineering the radiation pressure interaction of light with mechanical systems, comes as a powerful and well-developed tool to do so. First, radiation pressure interaction can be exploited to cool a (nano)micromechanical oscillator to its motional ground-state \cite{Genes1}; this is a necessary step for quantum manipulation and could not be accomplished by direct means such as cryogenic cooling (at the typical mechanical frequencies involved $100 KHz \sim 1 GHz$ this would require cooling the environment to a temperature of the order $1 \mu K \sim 10 mK$). Backaction cooling has been experimentally demonstrated for a variety of physical implementations, including micromirrors in Fabry-Perot cavities \cite{Groblacher1}, microtoroidal cavities \cite{Verhagen1} or optomechanical crystals \cite{Chan1}. Second, there exists a strong analogy between quantum optomechanics and non-linear quantum optics, so that many (if not all) optomechanical effects can be mapped onto well-known optical effects. As a result, optomechanics becomes a natural way for controlling a mechanical resonator at the quantum level. Experimentally, the strong coupling regime needed to observe quantum behaviors has been demonstrated only very recently \cite{Verhagen1, OConnel1}, and detection of quantum effects is still awaiting. Nevertheless, a lot of theoretical studies on the subject has been carried out in the last decade and several proposals have been produced \cite{Genes2}. These cover, among other things, the generation of entanglement between one oscillator and the radiation in a Fabry-Perot cavity \cite{Vitali1}, the generation of entanglement between two oscillators \cite{Vitali2}, or the generation of squeezed mechanical states \cite{Jahne1, Mari1}.
In particular,  references \cite{Mari1, Mari2, Galve1}
 introduced a new and effective way of enhancing the generation of quantum effects, which relies on applying a periodic modulation to some of the system parameters (a similar result has also been found in the analogous contest of nanoresonators and microwave cavities \cite{Woolley1}).

In this paper we further investigate the properties of periodically modulated optomechanical systems and we address the following questions: which is the fundamental link between modulation and enhancement of quantum effects? is there an optimal choice of the modulation, for which the visibility of quantum effects is maximal? Is this optimal regime robust against parameter fluctuations? What happens when two independent modulations are applied simultaneously?
To tackle these issues we analyze  the paradigmatic case of a mechanical  oscillator whose natural frequency 
$\omega_M$ is externally modulated when it evolves under the action of
the noise and of the radiation pressure exerted by the photons of an externally driven optical cavity mode.  While quantum optomechanics is nowadays extensively studied within a variety of experimental setups, the modulation of the mechanical frequency we analyze here is a very crucial aspect of our system and one that has not been implemented yet. However, very recent proposals for doing optomechanics with levitated dielectric spheres \cite{Chang1, Romero1, Li1} can be a good answer. In these proposals the mechanical degree of freedom is represented by the center of mass motion of a nanodielectric sphere which is trapped and levitated by means of an optical trap. The sphere is then put inside an optical cavity, where it interacts with the intracavity radiation via the usual optomechanical Hamiltonian (\ref{EqDimHam}). The frequency of the center of mass motion depends on the shape of the trapping potential and can thus be modulated adjusting the intensity of the trapping laser, as shown in \cite{Chang1}. Moreover, typical parameters attainable with such setups are comparable to those we have adopted in our simulations (see below), assuring the feasibility of the system under analysis.

In the above scenario we study the formation of  squeezing, entanglement and discord~\cite{DISCORD}, showing that in the steady state all these quantum effects  are  enhanced  when the modulation frequency $\Omega$ is twice the original value of $\omega_M$. As we shall see, such resonance admits a simple interpretation
in terms of an effective  parametric  phase-locking between 
the external driving forces and the natural 
evolution of the involved degree of freedom.
Similar enhancements  were also observed in  Refs. \cite{Mari1, Mari2}, where an harmonic modulation of  $2 \omega_M$ was imposed on the amplitude of the cavity mode laser, and in Ref. \cite{Galve1}, where a harmonic modulation of  $2 \omega_M$ was imposed on the coupling rate between two generic bosonic modes. Since several mechanisms can lead independently to the same effect, an interesting question is how they can be best exploited to control specific quantum properties in the system. This goes in the direction of developing optimal quantum control protocols, a topic which is currently benefiting from many contributions \cite{Rogers1}.
In the present case, to study the interplay of different mechanisms we add a second modulation in our model  and we observe the arising of interference pattern in the system response. Specifically we notice that the ability in cooling and  squeezing the mechanical oscillator  strongly depends upon the relative phase of the two modulations, the relative variation being almost 50$\%$. 

The material  is organized as follows. In section \ref{SecSys} we present the system and solve its dynamical evolution under the action of a  periodic modulation of the mechanical frequency. In section \ref{SecRes1} we then characterize the asymptotic stationary state in terms of entanglement, squeezing, etc. In section \ref{SecRes2} we compare our findings to other recent proposals \cite{Mari1, Mari2} and we study what happens when a second  independent modulation is applied to the system [specifically, in our case we introduce a modulation on the  amplitude of the input laser]. Conclusions and general remarks follows in section \ref{SecCon}. Some technical derivations are finally reported in Appendix~\ref{appa}.

\begin{figure}[t]
\includegraphics[width=0.45\textwidth]{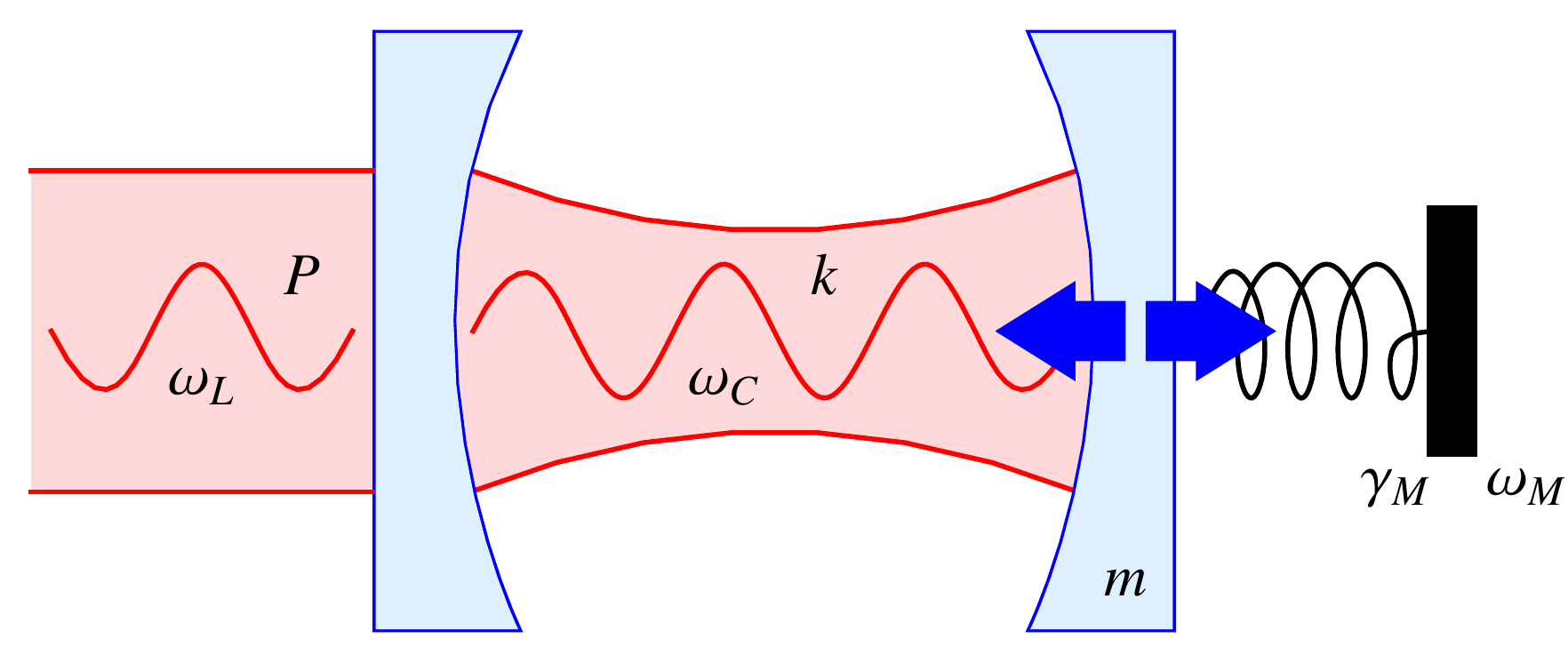}
\caption{Schematic description of the system. A Fabry-Perot cavity is driven by an external laser and the radiation interacts with the movable mirror on the right, exchanging momentum.}
\label{FigSys}
\end{figure}

\section{The system}
\label{SecSys}
Our choice falls on the simplest optomechanical system of all, i.e. a Fabry-Perot cavity of lenght $l_0$ with a movable mirror at one end (see Fig. \ref{FigSys}), which nevertheless captures all interesting physics. We can reasonably assume \cite{Genes2} that a single optical mode is interacting with a single mechanical mode, be it the center of mass oscillation. The mirror can thus be modeled as a mass $m$ attached to a spring of characteristic frequency $\omega_M$ and friction coefficient $\gamma_M$; it is described by dimensionless position and momentum operators $\hat q$, $\hat p$ which obey the canonical commutation relation $[\hat q, \hat p]=i$. The optical mode has frequency $\omega_C$ and decay rate $k$; it is described by annihilation/creation operators $\hat a$ and $\hat a^\dagger$, which obey the canonical commutation relation $[\hat a, \hat a^\dagger]=1$. 

In our analysis  the cavity is assumed to be driven by an external laser which, to begin with, we take to have constant power $P_\text{laser}$ and quasi-resonant frequency $\omega_L \sim \omega_C$. In this context a periodic modulation is inserted at the level of the  spring constant, which we express as the following time dependent parametric rescaling of the mirror frequency 
\begin{eqnarray}
\omega^2(t) = \omega_M^2\; [1 + \epsilon \cos (\Omega t)], \label{modulation}
\end{eqnarray}
with $\epsilon<1$. Accordingly the Hamiltonian of the system writes as \cite{Law1}
\begin{align}
	\hat{ \mathcal{H}} = & \hbar \omega_C \hat a^\dagger \hat a + \frac{\hbar \omega_M}{2} \hat p^2 + \frac{\hbar \omega_M}{2} \;[1 + \epsilon \cos (\Omega t)] \;\hat q^2 \notag \\
	& - \hbar G_0 \hat a^\dagger \hat a  \hat q + i \hbar E (e^{- i \omega_L t} \hat a^\dagger - e^{i \omega_L t} \hat a),
	\label{EqDimHam}
\end{align}
where $G_0 = \omega_C/l_0 \sqrt{\hbar/(m \omega_M)}$ is the optomechanical coupling rate and $|E| = \sqrt{2 k  P_\text{laser}  / \hbar \omega_L}$ is the driving rate.
Including dissipation and decoherence effects the system dynamics can then be described with the
 following set of quantum Langevin equations~\cite{Genes2} 
\begin{equation}
	\begin{cases}
	\partial_t \; \hat q = \omega_M \hat p,\\
	\partial_t \; \hat p = -\omega_M \;[1 + \epsilon \cos (\Omega t)]\; \hat q - \gamma_M \hat p + G_0 \hat a^\dagger \hat a + \hat \xi,\\
	\partial_t \; \hat a = -(k + i \Delta_0) \hat a + i G_0 \hat a \hat q + E + \sqrt{2k}\;  \hat a_{in},
	\end{cases}
	\label{EqSet}
\end{equation}
which we have written in a frame rotating at $\omega_L$.
Here $\Delta_0=\omega_C - \omega_L$ is the unperturbed cavity laser detuning while $\hat a_{in}(t)$ is the radiation vacuum input noise with autocorrelation function \cite{Gardiner1}
\begin{equation}
	\left< \hat a_{in}(t) \hat a_{in}^\dagger(t^\prime) \right> = \delta(t-t^\prime).
	\label{EqVacNoise}
\end{equation}
Similarly $\hat \xi(t)$ is the Brownian noise operator describing the dissipative friction forces acting on the mirror. Its autocorrelation function satisfies the relation~\cite{Giovannetti1}
\begin{equation}
	\left< \left\{ \hat \xi(t) ,\hat \xi(t^\prime)\right\}\right> = 2 \frac{\gamma_M}{\omega_M} \int \frac{d\omega}{2 \pi}  \; \omega \; \coth \left( \tfrac{\hbar \omega}{2 k_B T} \right) e^{-i \omega (t-t^\prime)}.
	\label{EqBrownNoise}
\end{equation}
which for the specific case of an harmonic oscillator with a good quality factor $\omega_M \gg \gamma_M$, 
acquires 
the same Markov character of Eq.~(\ref{EqVacNoise}), i.e.
\begin{equation}
	\left< \left\{ \hat \xi(t) ,\hat \xi(t^\prime)\right\}\right>  \approx 2
	 \gamma_M \coth \left( \tfrac{\hbar \omega_M}{2 k_B T} \right) \delta(t-t^\prime),
	\label{EqBrownNoiseMark}
\end{equation}
(this is a consequence of the fact that for $\omega_M \gg \gamma_M$ only resonant noise components at frequency $\omega \sim \omega_M$ do sensibly affect the motion of the system). 
In the above expressions $T$ is the system temperature while $\{ \cdots, \cdots\}$ is the anti-comutator~\cite{NOTAEXP}. 

\subsection{Solving the dynamics}
\label{SecSol}

The evolution of the system is ruled by a set (\ref{EqSet}) of non-linear stochastic differential equations with periodic coefficients, whose solution is in general very difficult. In the following we will then introduce some useful approximations to simplify the calculations. First, we expand each operator as the sum of a c-number mean value and a fluctuation operator, i.e. 
\begin{eqnarray}
\hat a(t) &=& \left< \hat a(t) \right> + ( \hat a(t) - \left< \hat a(t) \right> ) \equiv A(t) + \delta \hat a(t), \nonumber \\
\hat q(t) &\equiv&  Q(t) + \delta \hat q(t), \nonumber \\
\hat p(t) &\equiv & P(t) + \delta \hat p(t).
\end{eqnarray}
 We recall that the cavity is usually driven by a very strong laser in order to attain satisfactory levels of optomechanical interaction, so that the mean value will be much bigger than the fluctuations, which are due to the presence of random noise. This allows us to write (\ref{EqSet}) as two different sets of equations, one for the mean values (\ref{EqSetMean}), one for the fluctuations (\ref{EqSetFluc}) and linearize the latter neglecting all terms which are second order small, obtaining 
\begin{equation}
	\begin{cases}
	\partial_t \; Q = \omega_M P,\\
	\partial_t \; P = -\omega_M \;[1 + \epsilon \cos (\Omega t)]\; Q - \gamma_M P + G_0 |A|^2,\\
	\partial_t \; A = -(k + i \Delta_0) A + i G_0 A Q + E,
	\end{cases}
	\label{EqSetMean}
\end{equation}
\begin{widetext}
\begin{equation}
	 \partial_t \left( \begin{array}{c} \delta \hat q \\ \delta\hat p \\ \delta\hat X \\ \delta \hat Y \end{array} \right) =  \left( \begin{array}{cccc} 0 & \omega_M & 0 & 0 \\ -\omega_M \;[1 + \epsilon \cos (\Omega t)]\;& -\gamma_M & G_0 Re[A] & G_0 Im[A] \\ -G_0 Im[A] & 0 & -k & \Delta_0 - G_0 Q \\ G_0 Re[A] & 0 & -\Delta_0 + G_0 Q & -k \end{array} \right) \cdot \left( \begin{array}{c} \delta \hat q \\ \delta\hat p \\ \delta\hat X \\ \delta \hat Y \end{array} \right) + \left( \begin{array}{c} 0 \\ \hat \xi \\ \hat X_{in} \\ \hat Y_{in} \end{array} \right),
	\label{EqSetFluc}
\end{equation}
\end{widetext}
where we have introduced the phase and amplitude quadratures for the cavity and the input noise fields, i.e.  \linebreak $\hat X = (\hat a^\dagger + \hat a)/\sqrt{2}$, $\hat Y = i(\hat a^\dagger - \hat a)/\sqrt{2}$, $\hat X_{in} = (\hat a_{in}^\dagger + \hat a_{in})/\sqrt{2}$ and $\hat Y_{in} = i(\hat a_{in}^\dagger - \hat a_{in})/\sqrt{2}$.
 Equation~(\ref{EqSetFluc}) can be also expressed in a more compact form 
\begin{equation}
	\partial_t \; \hat u = S \hat u + \hat \zeta,
	\label{EqComp}
\end{equation}
with $S$ being a $4\times 4$ time-dependent matrix, and 
with $\hat{u}$ and $\hat{\zeta}$ being the column vectors of elements $(\delta \hat q, \delta\hat p, \delta\hat X , \delta \hat Y)$ and $(0, \hat \xi, \hat X_{in}, \hat Y_{in})$, respectively. 
We stress that Eqs.~(\ref{EqSetMean}) and (\ref{EqSetFluc}) must be solved in the correct order, because the mean values $Q(t)$, $P(t)$ and $A(t)$ play the role of coefficients in the equations for the fluctuations.

\begin{figure}[t]
\includegraphics[width=0.45\textwidth]{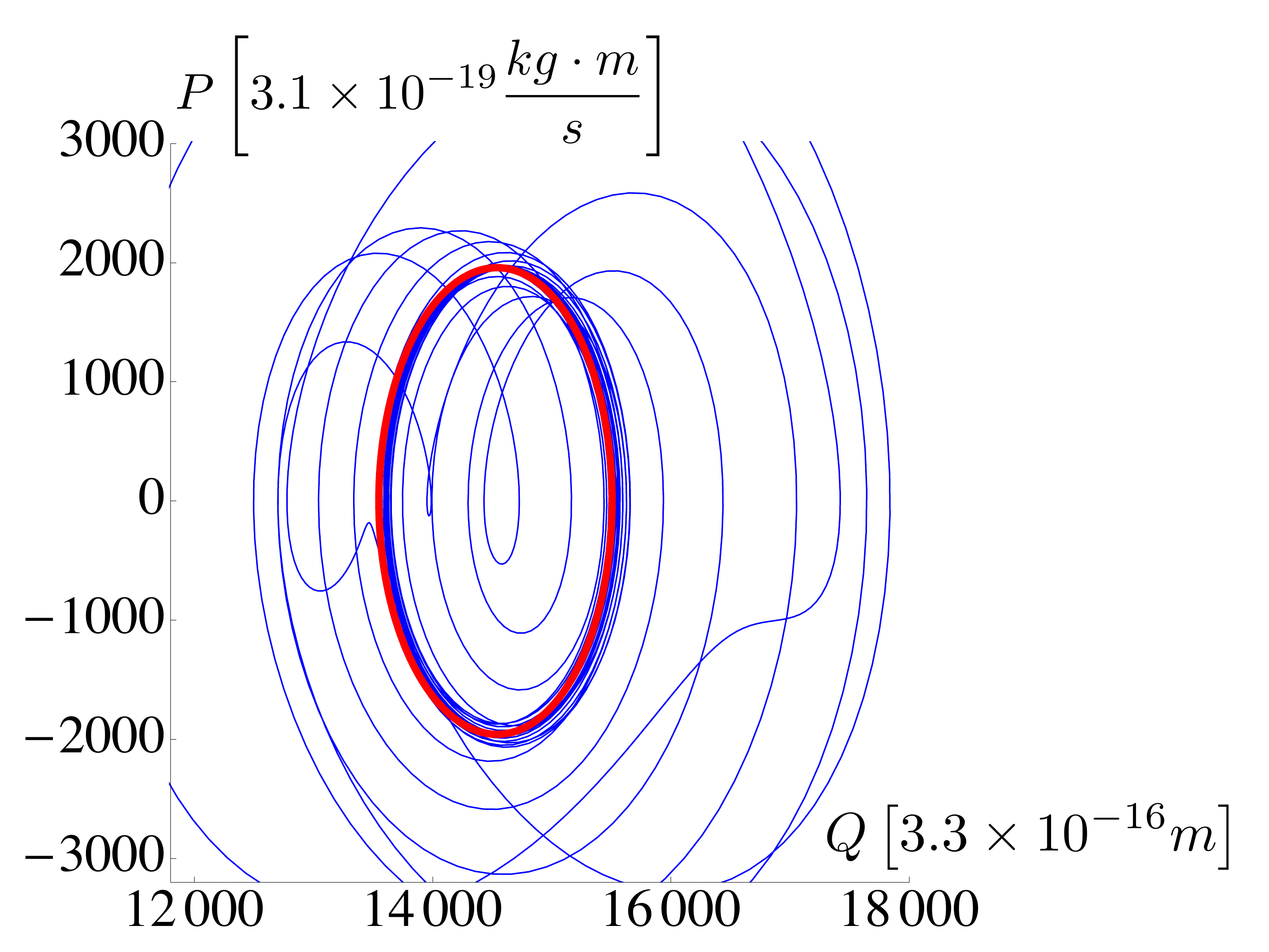}
\caption{Evolution of the mirror position $Q(t)$ and momentum mean  values $P(t)$   obtained by numerically integrating Eq.~(\ref{EqSetMean}) from $t=0$ to $t= 50 \tau$, with $\tau=2\pi/\Omega$ being the
period of the modulation (thin blue line). 
The plot has been obtained by setting the system parameters as detailed in 
Sec.~\ref{SecRes1}: in particular here the modulation frequency $\Omega$ is twice the
natural frequency  $\omega_M$ of the mechanical oscillator which, in turn, is resonant with the detuning 
$\Delta_0$ that  governs the free evolution of the optical field $A(t)$. The analytic solution for the asymptotic orbit (see Appendix~\ref{AppSingleMod}) is also shown  for comparison (thick red line).}\label{FigMeanQP}
\end{figure}

Equation~(\ref{EqSetMean}) is nonlinear but  can be solved numerically.
Assuming that we are far from optomechanical instabilities and that  we keep the modulation strength $\epsilon$ small enough to avoid additional instabilities due to parametric amplification,
 one finds that the mean values evolve toward an asymptotic periodic orbit with the same periodicity $2\pi / \Omega$ of the applied modulation. 
In this regime, an approximate analytic solution can also be derived, which we detail in Appendix~\ref{appa}. Indeed since the modulation strength $\epsilon$ is not too strong, one can guess a perturbative expansion of the form 
\begin{eqnarray}
Q(t) = \sum\limits_{j=0}^\infty Q^{(j)}(t)\;, \label{ANapprox1}
\end{eqnarray}
 where $Q^{(0)}(t)$ does not depend on $\epsilon$, $Q^{(1)}(t)$ is linear in $\epsilon$, $Q^{(2)}(t)$ is quadratic in $\epsilon$ and so on. It turns out that each order is exactly solvable, as long as previous orders are known. This originates a chained set of equations and by keeping a finite number of orders $j \leq j_{MAX}$, we can finally obtain the asymptotic solution up to the desired precision (e.g. see Fig. \ref{FigMeanQP}).

Equation (\ref{EqSetFluc}) is stochastic and needs some more manipulation. Nonetheless since we have linearized the dynamics and the noises are zero-mean gaussian noises, fluctuations in the stable regime will also evolve to an asymptotic zero-mean gaussian state. The state of the system is then completely described by the correlation matrix ${C}$ of elements
\begin{equation}
	{C}_{ij}(t) = {C}_{ji}(t)=\frac{1}{2} \left< \hat u_i(t) \hat u_j(t) + \hat u_j(t) \hat u_i(t) \right>,
	\label{EqCorrMat}
\end{equation}
whose evolution can be derived directly from equations (\ref{EqComp}) and (\ref{EqCorrMat}):
\begin{equation}
	\partial_t {C} = S {C} + {C} S^\top + N,
	\label{EqComp2}
\end{equation}
where $S^{\top}$ is transpose of $S$, and where $N$ is the diagonal noise correlation matrix with diagonal entries $(0,\gamma_M \coth(\hbar \omega_M / 2 k_B T),k,k)$, defined by 
\begin{eqnarray}
\frac{1}{2} \left< \hat \zeta_i(t) \hat \zeta_j(t') + \hat \zeta_j(t') \hat \zeta_i(t) \right> \equiv N_{ij} \delta(t-t').
\end{eqnarray} 
Equation (\ref{EqComp2}) is now an ordinary linear differential equation.
We know that its solution evolves toward a unique asymptotic configuration (independently of the initial state), proven that the eigenvalues of the matrix $S$ have negative real part for all times $t$, which can be verified by applying the Routh-Hurwitz criterion \cite{DeJesus1}. Again, we can either solve Eq.~(\ref{EqComp2}) numerically or obtain an approximate analytic solution with a perturbative expansion in $\epsilon$ (see Appendix~\ref{appa} for the latter).


\subsection{Quantum properties of the system}
\label{SecQuantumProp}
As already mentioned, thanks to gaussianity of the asymptotic solution all relevant informations about the system can be extracted directly from the correlation matrix ${C}$. 
In particular we will focus on the following quantities: the number of phonons in the mirror, the squeezing in the mirror and in the radiation quadratures, and the nonclassical correlation between the mirror and the radiation degrees of freedom.

The number of phonons $n$ can be expressed using the approximate relation
\begin{eqnarray}
	\hbar \omega_M \left(n + \tfrac{1}{2} \right) &\approx& \hbar \omega_M/2 \; \left< \delta q^2 + \delta p^2 \right> 
	\nonumber \\ &=& 
	 \hbar \omega_M/2 \left( {C}_{11} + {C}_{12} \right),
	\label{EqPhonons}
\end{eqnarray}
which holds if the modulation of the mechanical frequency is not too strong. This tells how far the system is from the ground state. Since both ${C}_{11}$ and ${C}_{12}$ are periodic in time, we will identify the number of phonons with the maximum over one period $\tau = 2 \pi / \Omega$ of the modulation, i.e. 
\begin{equation}
	n_{MAX} = \max_{\tau} \left\{ n(t) \right\},
	\label{EqMaxPhonons}
\end{equation}
 (here and in the following $\max_{\tau}$ represents an optimization with respect to a time interval 
$[{\cal T}, {\cal T} + \tau]$ with ${\cal T}$ being sufficiently larger than $1/k$ to guarantee that the system
has reached the asymptotic steady state).

Squeezing of the generalized mirror quadratures $q_{\theta} = q  \cos\theta + p  \sin\theta$ is also easily found:
\begin{equation}
	\left< \delta q^2_{\theta} \right> = {C}_{11} \cos^2\theta + {C}_{22} \sin^2\theta +  ( {C}_{12} + {C}_{21} ) \cos\theta \sin\theta.
	\label{EqVariance}
\end{equation}
Again we construct a time independent quantity to deal with. First, for each time $t$ we select the parameter $\theta$ for which $\left< \delta q^2_{\theta} \right>$ is minimum. In terms of covariance matrix, this is just the smaller eigenvalue of the block matrix $\left( \begin{array}{cc} {C}_{11} & {C}_{12} \\ {C}_{21} & {C}_{22} \end{array} \right) $. We then minimize this quantity with respect to time over a period $\tau$. This tells how much squeezing can be produced at most.
\begin{equation}
	\Delta^2 q_{MIN} = \min_{\tau} \left\{  \min_{\theta} \left< \delta q^2_{\theta} \right>  \right\}.
	\label{EqMinMirrorVariance}
\end{equation}
Analogous formulas for the radiation quadratures lead to
\begin{equation}
	\Delta^2 X_{MIN} = \min_{\tau} \left\{  \min_{\theta} \left< \delta X^2_{\theta} \right>  \right\}.
	\label{EqMinRadiationVariance}
\end{equation}
Non-classical correlations in the system can be described using quantum discord $\mathcal{D} (\rho)$~\cite{DISCORD}, which includes entanglement as well as more general quantum correlations that are shown also by separable states \cite{Modi1}. For a gaussian state, $\mathcal{D} (\rho)$ is easily constructed from the correlation matrix as demonstrated in \cite{Adesso1}. Time dependance is then eliminated by considering
\begin{equation}
	\mathcal{D}_{MAX} = \max_{\tau} \left\{ \mathcal{D} (\rho(t))  \right\}.
	\label{EqMaxDiscord}
\end{equation}
Entanglement alone will be specifically described using logarithmic negativity $E_{\mathcal{N}} (\rho)$ \cite{Vidal1}, which is also easily constructed from the correlation matrix as demonstrated in \cite{Ferraro1}. Again, time dependance is eliminated by considering
\begin{equation}
	E_{\mathcal{N}_{MAX}} = \max_{\tau} \left\{ E_{\mathcal{N}} (\rho(t))  \right\}.
	\label{EqMaxEntanglement}
\end{equation}

\section{Results}
\label{SecRes1}
We now present the results  obtained by solving 
the dynamics of the system as detailed in the previous section. The parameters used in our analysis are $m = 150$ ng, $\omega_M/(2\pi) = 1$ MHz, $\gamma_M/(2 \pi) = 1$ Hz, $T=0.1$ K, $\Delta_0 = \omega_M$, $l_0 = 25$ mm, $k = 1.34$ MHz, $\lambda = 1064$ nm and $P_\text{laser}  =10$ mW: this choice is compatible with values attained in state of the art experiments and is also consistent with the stability requirement of section \ref{SecSol}  (furthermore, under the condition $\Delta_0 = \omega_M$
the optical and the mechanical variables are brought at resonance).
The strength $\epsilon$ and the frequency $\Omega$ of the modulation are left as variable parameters instead, since we want to characterize the optimal modulation regime, e.g. which $\epsilon$ and $\Omega$ maximize the visibility of quantum effects.\\

In Fig. \ref{FigMeanQP}, we temporarily fix  $\Omega = 2 \omega_M$, $\epsilon=0.2$ (this particular choice will be justified in the following) and we report the solution of Eq~(\ref{EqSetMean}) for the mean values $Q(t)$ and $P(t)$ of the mirror position and momentum. We see that the evolution tends indeed to an asymptotic periodic orbit, which is very well approximated by the analytic solution.\\

\begin{figure*}[ht!]
\centering
\begin{minipage}{0.95\textwidth}
\includegraphics[width=0.45 \textwidth]{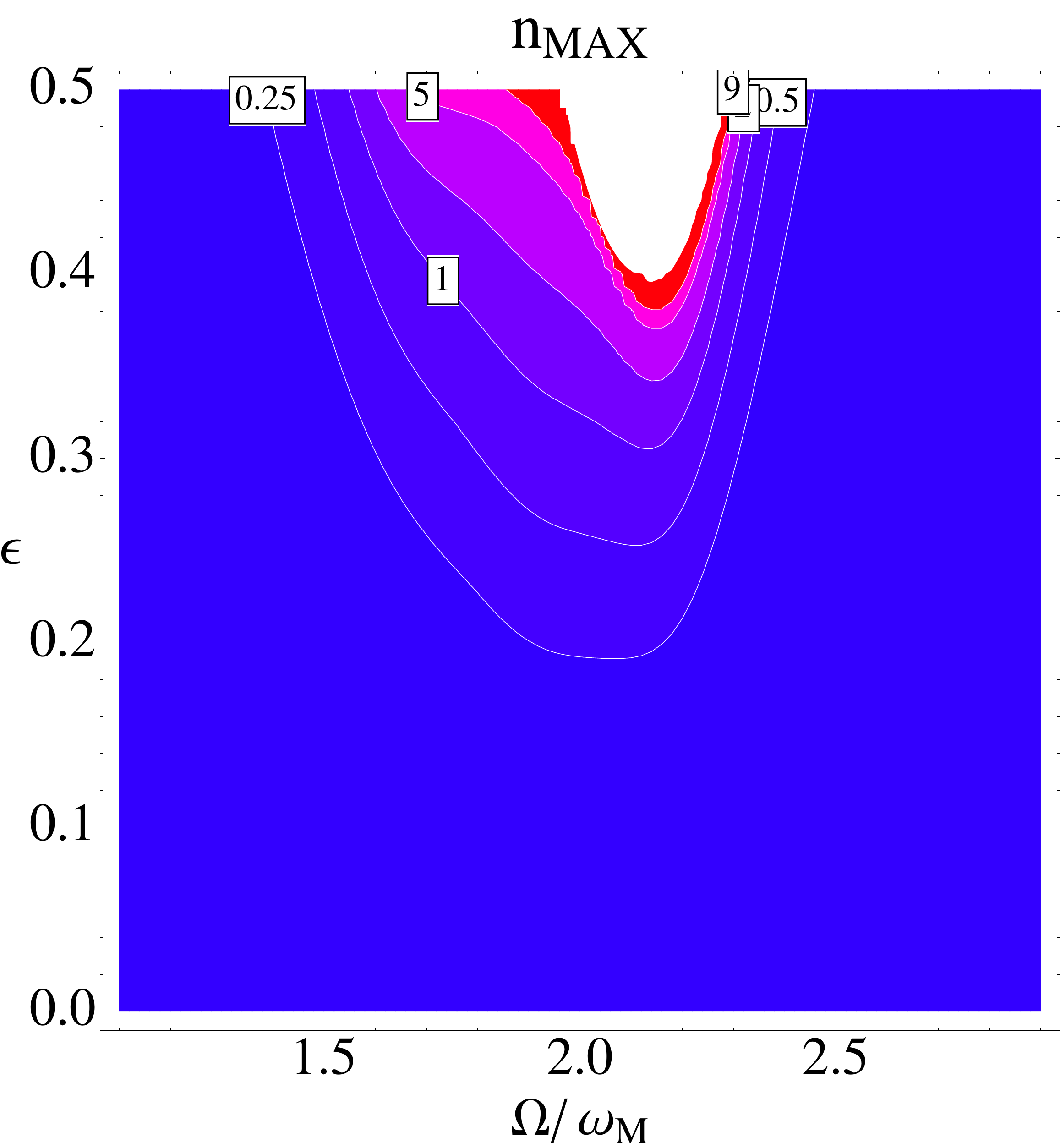}\hspace{0.04\textwidth}
\includegraphics[width=0.45 \textwidth]{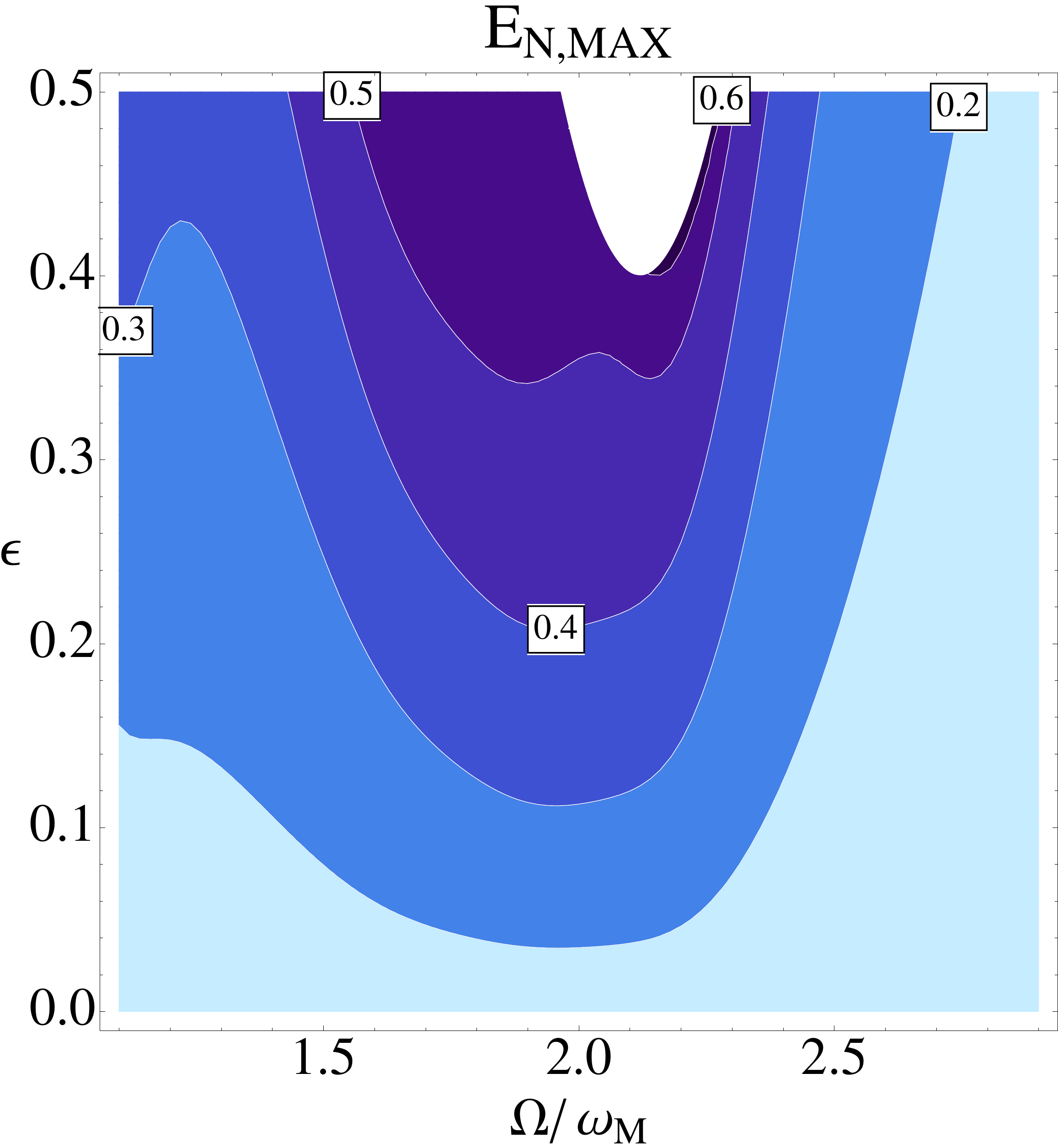}
\includegraphics[width=0.45 \textwidth]{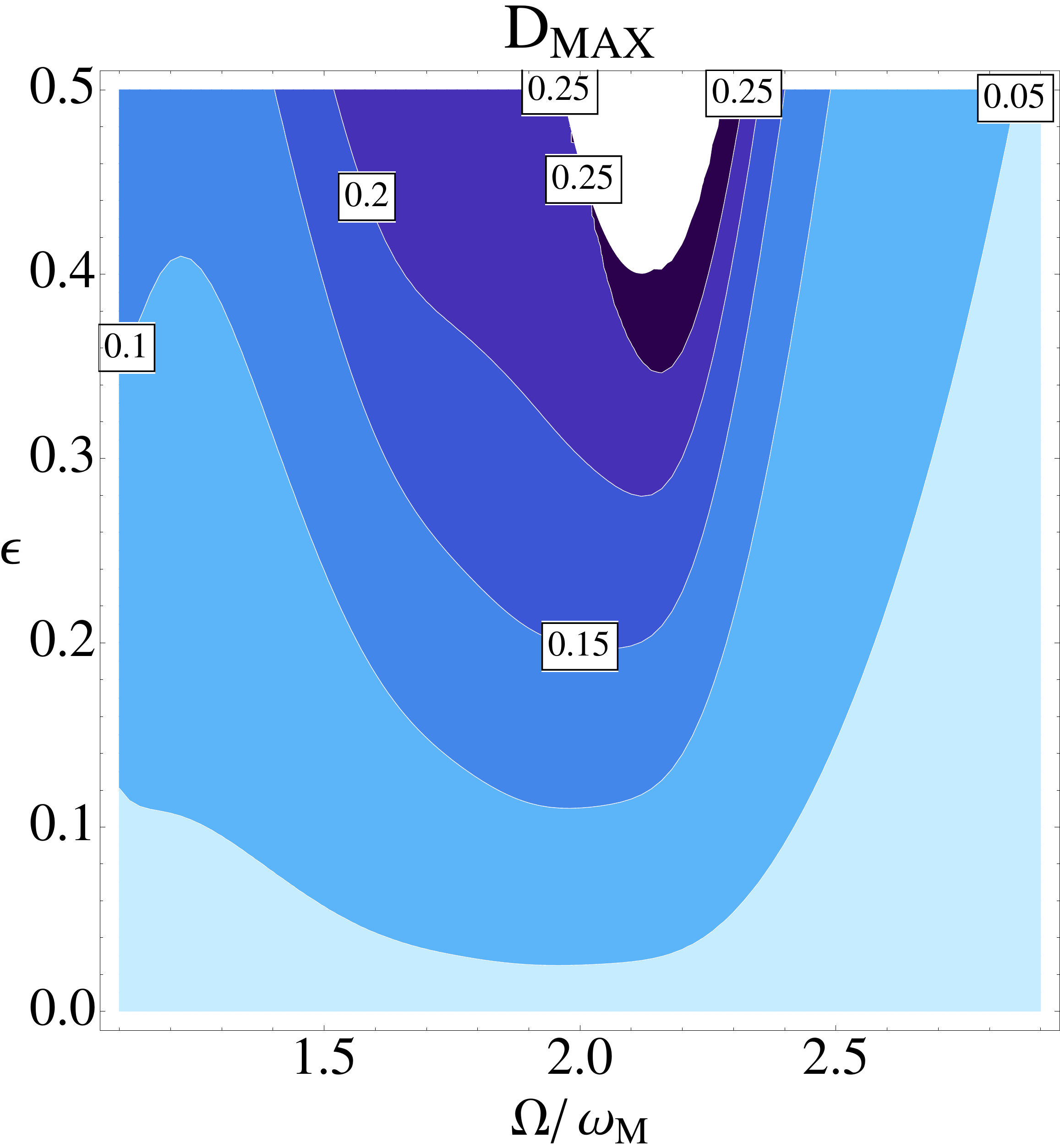}\hspace{0.04\textwidth}
\includegraphics[width=0.45 \textwidth]{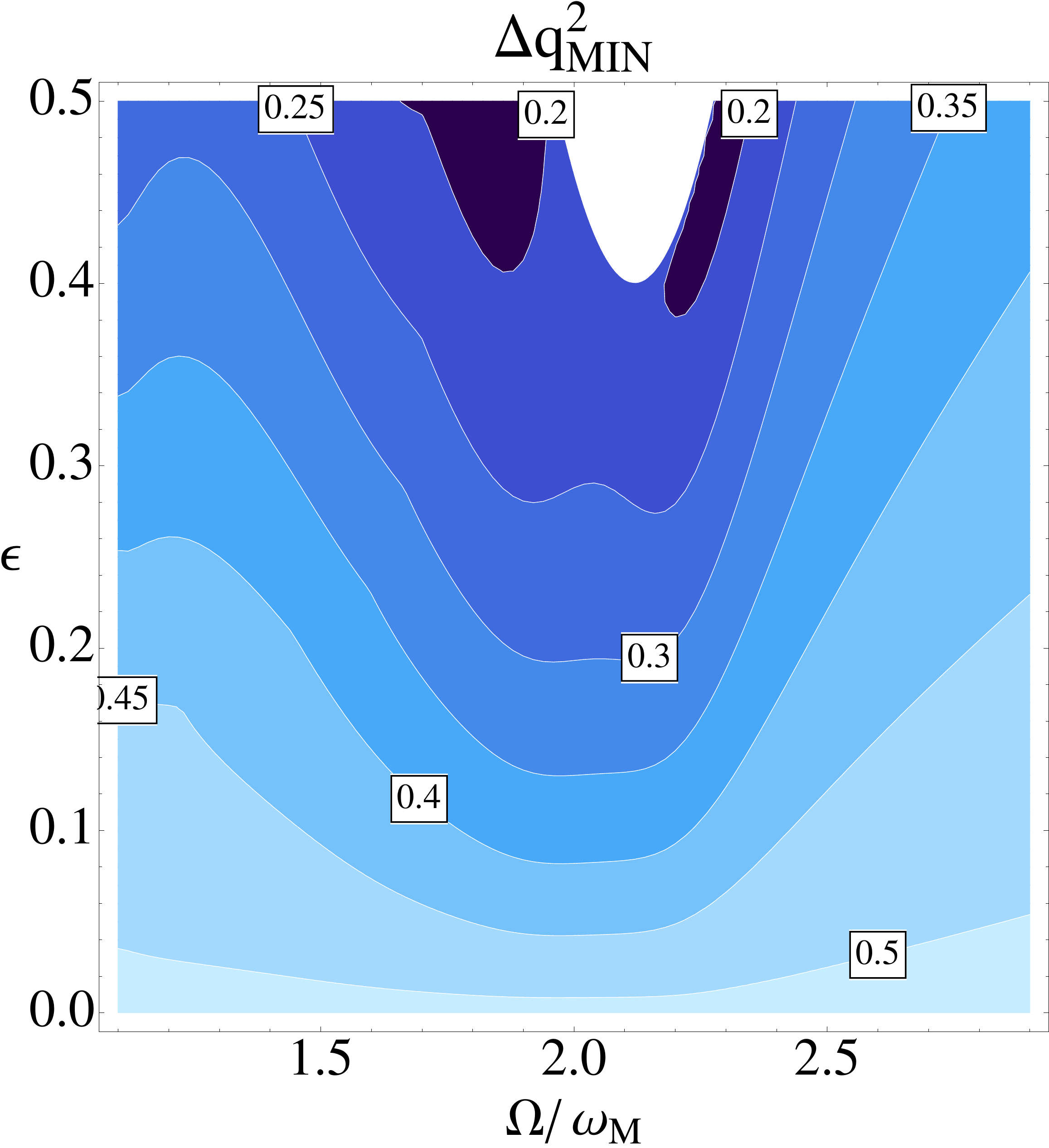}
\caption{Asymptotic quantum features as a 
 function of  $\Omega/\omega_M$ (x axis) and $\epsilon$ (y axis):
 a) Maximum number of phonons in the mirror, eq (\ref{EqMaxPhonons}); 
 b) Maximum of the logarithmic negativity (\ref{EqMaxEntanglement});
 c) Maximum of the quantum discord (\ref{EqMaxDiscord});
 d) Minimum of the generalized quadratures of the mirror (\ref{EqMinMirrorVariance}).
In all the plots the
 system parameters are fixed as in Sec.~\ref{SecRes1}. }
		\label{FigMaxPhonons}
		\end{minipage}
\end{figure*}

We then focus on the solution of equation (\ref{EqComp2}) and we plot the quantities described in section \ref{SecQuantumProp}, for multiple values of  $\Omega \in [\omega_M, 3 \omega_M]$, $\epsilon \in [0,0.5]$.
In particular: Fig.~\ref{FigMaxPhonons} shows the maximum number $n_{MAX}$ of phonons in the mirror, computed via Eq.~(\ref{EqMaxPhonons}); 
 the maximum $E_{\mathcal{N}_{MAX}}$ of the logarithmic negativity, computed via Eq.~(\ref{EqMaxEntanglement});
the maximum $\mathcal{D}_{MAX}$ of the quantum discord, computed via Eq.~(\ref{EqMaxDiscord}); 
and the minimum variance $\Delta^2 q_{MIN}$ of all the mirror generalized quadratures, computed via Eq.~(\ref{EqMinMirrorVariance}).

As evident from the plots, the level of squeezing and entanglement is maximum when the modulation frequency is $\Omega \sim 2\omega_M$ and increases monotonically with respect to the strength $\epsilon$, until the system eventually reaches an instability point for too strong modulations (in the above figures, this instability is represented by a blank region around the point $\epsilon=0.5$, $\Omega \sim 2 \omega_M$). It is also clear that the optimal modulation, the one that most enhances quantum effects, is also responsible for heating the system far from its ground state.
We can understand this behavior if we interpret Eq.~(\ref{EqComp2})  as describing the dynamics
of a set of (classical) parametric oscillators with canonical coordinates defined by the correlations functions 
${C}_{ij}$ (\ref{EqCorrMat}), which evolve under the action of damping and constant external driving 
forces. Indeed, by a close inspection of the matrix $S$ 
one notices that such oscillators possess natural frequencies which are periodically modulated 
 through functions   (i.e. $A(t)$, $Q(t)$, and the direct term $\epsilon \cos (\Omega t)$) that, in first approximation, 
 evolve sinusoidally with the same frequency $\Omega$ -- see Eq.~(\ref{EqQAnalytic1Mod}) in Appendix~\ref{appa} for details. Moreover, in the stability region we are sure that parametric modulation pumps energy into the system at a lower rate with respect to losses, since the system evolves toward a stationary orbit: we call this regime ``below-threshold'' to distinguish it from the exponential amplification usually associated with parametric oscillators.
 For this model phase-locking is expected to occur when $\Omega$ matches the zero-order eigenfrequencies defined by
 the constant part of $S$ (and not twice this frequencies as in the case of parametric instability), resulting in an enhancement of the oscillations of the effective coordinates 
${C}_{ij}$ (\ref{EqCorrMat}) and hence of the associated quantum effects defined in Sec.~\ref{SecQuantumProp}~\cite{NOTE111} (more details are found in Appendix~\ref{appa}).
It turns out that, at least for the figure of merit we are concerned here (i.e. $\Delta^2 q_{MIN}$, $\Delta^2 X_{MIN}$, $\mathcal{D}_{MAX}$, etc)
 the relevant frequency is indeed $\sim 2\omega_M$.
 
To see this, we can procede by steps. First of all notice that 
from the numerical solution, we can guarantee  that the system is not unstable (see Fig.~\ref{FigMaxPhonons}), i.e. that it is indeed in the  below-threshold regime. 
Next, consider the case of no coupling ($G_0 = 0$) and no modulation ($\epsilon =0$): we stress out three relevant aspects. First, the mechanical part and the radiation part are independent, so there is no entanglement. Second, each subsystem evolves with the Hamiltonian of a quantum harmonic oscillator, so the quadrature mean value $\left< q_\theta \right>$ evolves with a phase $e^{i \omega_M t}$ and the variance $\left< \delta q_\theta^2 \right>$ with a phase $e^{i 2 \omega_M t}$ (we remind that we fixed $\Delta_0=\omega_M$). This tells us that, at least in this regime,  the 
frequencies which govern the quantities of interest are degenerate  at the  value $2 \omega_M$. Third, each subsystem is also coupled to its own environment and will
eventually relax to a thermal state characterized by $\left< \delta q_\theta^2 \right> = N_{therm} + 1/2$, so there is no squeezing.
Now turn on the coupling $G_0$: this has three main effects. First it introduces  entanglement in  the system \cite{Vitali1} ($E_{\mathcal{N}} = E_0$). 
Second the eigenfrequencies are brought out of degeneracy and shifted by a term $\propto 2 G_0 \left| A \right|$ \cite{Dobrint1}, which is quite small with respect to $2 \omega_M$ for our choice of values
(confirming that
indeed the latter is the resonant value at which the modulation should provide an enhancement). Third, 
backaction cooling \cite{Genes1} is now active and the oscillator approaches the ground state $\left( \left < \delta q_\theta^2 \right> \sim 1/2 \right)$. Squeezing is still absent at this level. 
Finally, turn on the modulation~(\ref{modulation}). Thanks to the phase-locking mechanism we have anticipated previously and detailed in Appendix~\ref{appa}
this will yield an enhancement of the correlations when $\Omega$ matches the natural frequency $\sim 2 \omega_M$. For instance, for the negative entropy $E_{\mathcal{N}}$
and for  mirror variance $\left<  \delta q_\theta^2 \right>$, we get 
\begin{eqnarray}
	E_{\mathcal{N}} &\sim& E_0 + \epsilon\; K_1(\Omega) \; \cos (\Omega t + \varphi_1), \nonumber \\
	\left<  \delta q_\theta^2 \right>_0 &\sim& 1/2 + \epsilon\;K_2(\Omega) \; \cos (\Omega t + \varphi_2), \label{NEW}
\end{eqnarray}
where $K_1(\Omega)$ and $K_2(\Omega)$ are associated response functions
analogous to the Lorentzian response of a simple harmonic oscillator~(though an exact expression is rather cumbersome in our specific case) and are peaked around $\Omega \sim 2 \omega_M$. We see that the quadrature $\delta q_\theta^2$ gets periodically squeezed over time and entanglement is periodically increased to higher values with respect to the unmodulated case.
In addition, these effects increase monotonically with $\epsilon$, up to the instability threshold. 
 A similar enhancement of the entanglement is also described in Ref. \cite{Galve1}, where two harmonic oscillators are coupled via linear interaction $H_{int} = c(t) X_1 X_2$ and the coupling constant is a periodic function of time. This time dependance produces an effective modulation on the normal frequencies of the system: as a result, entanglement is shown to increase and become much more robust against temperature. This agrees very well with what we found here.

\begin{figure*}[t]
\centering
\begin{minipage}{0.95\textwidth}
\includegraphics[width=0.45 \textwidth]{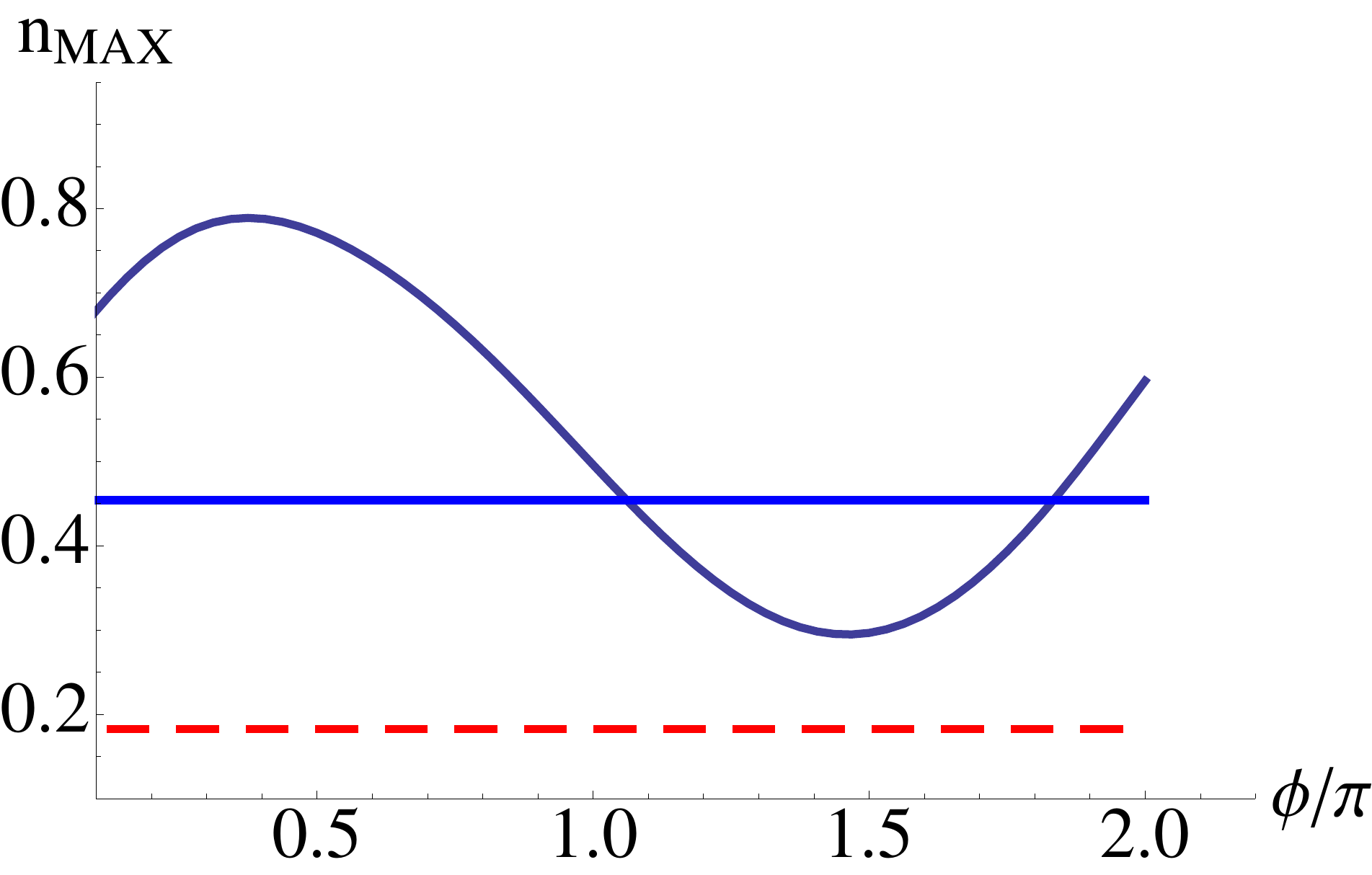}\hspace{0.04\textwidth}
\includegraphics[width=0.45 \textwidth]{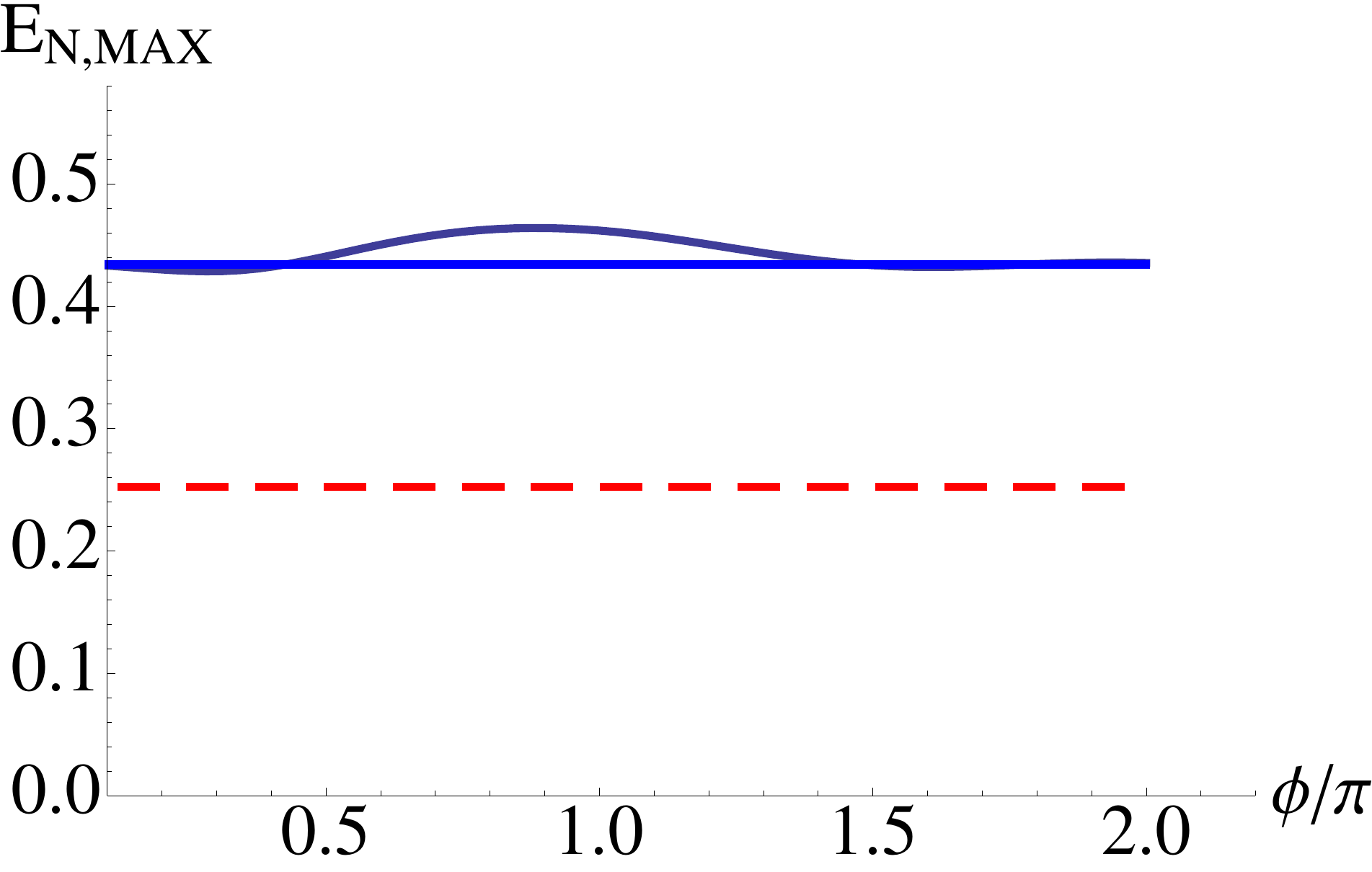}
\includegraphics[width=0.45 \textwidth]{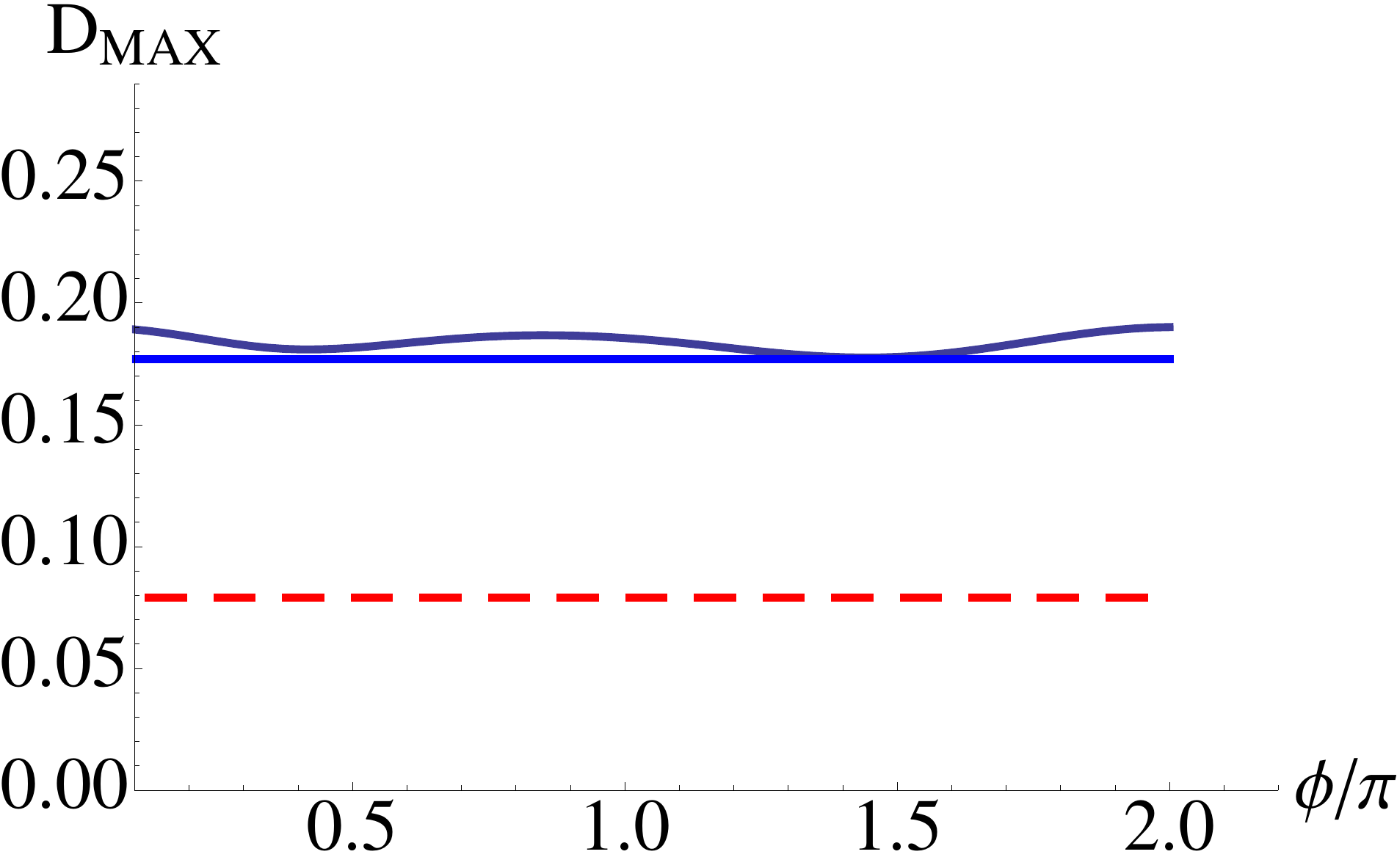}\hspace{0.04\textwidth}
\includegraphics[width=0.45 \textwidth]{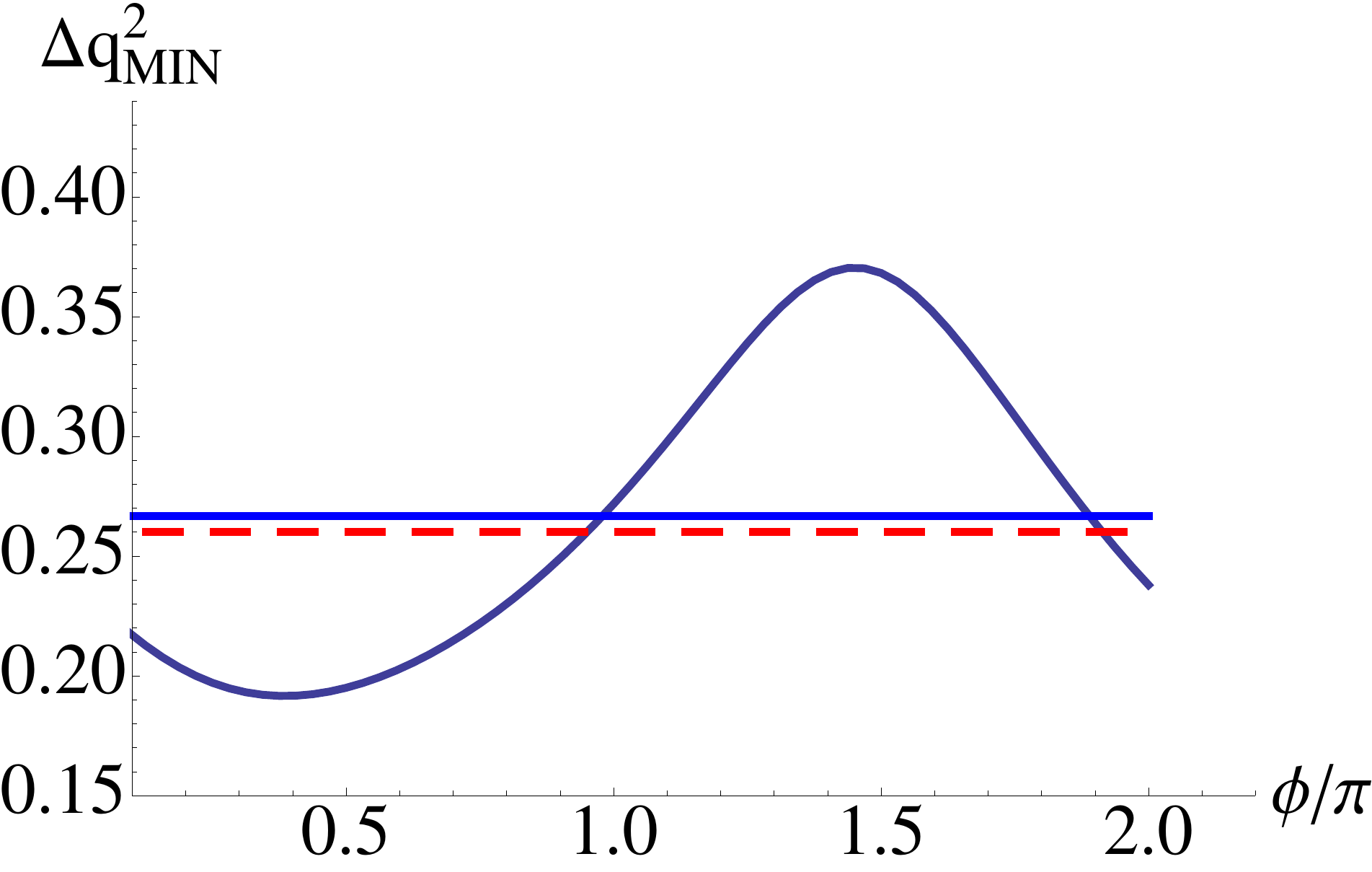}
\caption{Response of the system in the presence of two different modulations as a function 
of their relative phase $\phi$:
a) Maximum number of phonons in the mirror;
b) Maximum of the logarithmic negativity;
c) Maximum of the quantum discord;
d) Minimum of the generalized quadratures of the mirror.
 In all the plots the two straight lines show the variation of the function  in the case when only the mechanical frequency (blue horizontal) or the laser amplitude (red dashed horizontal) is modulated. Parameters as detailed in  the text. }
\label{FigTwoModPhonons}
	\end{minipage}
\end{figure*}

\section{Interplay between two different modulations}
\label{SecRes2}
 Results analogous to those presented in the previous section  has been found very recently by Mari and Eisert \cite{Mari1, Mari2}, for an optomechanical system driven with an amplitude modulated input laser. For clarity, we rewrite their Hamiltonian
\begin{align}
	\hat{ \mathcal{H}} = & \hbar \omega_C \hat a^\dagger \hat a + \frac{\hbar \omega_M}{2} \big( \hat p^2 + \hat q^2 \big) - \hbar G_0 \hat a^\dagger \hat a  \hat q   \notag \\
	& + i \hbar \big(E + E_1 \cos(\Omega t) \big) (e^{- i \omega_L t} \hat a^\dagger - e^{i \omega_L t} \hat a).
	\label{EqMariHam}
\end{align}
At first sight, the situation appears to be somewhat different from our initial problem. In Eq.~(\ref{EqMariHam}), internal parameters of the system are left unchanged; it is instead the external driving that undergoes an oscillatory behavior. Nevertheless the effects are strikingly similar: high levels of squeezing can be attained when the frequency of modulation is $\Omega \sim 2 \omega_M$ \cite{Mari1}, and the same regime is also optimal to enhance entanglement between mechanical and radiation modes \cite{Mari2} The authors themselves comment that \emph{``...this dynamics reminds of the effect of parametric amplification, as if the spring constant of the mechanical motion was varied in time with just twice the frequency of the mechanical motion, leading to the squeezing of the mechanical mode..."} \cite{Mari1}.

In fact there is a strong analogy between the two cases. Independently of what Hamiltonian (\ref{EqDimHam}) or (\ref{EqMariHam}) one chooses, far from instability regions the mean values $Q(t)$, $P(t)$, $A(t)$ will be characterized by an asymptotic periodic orbit with the same periodicity of the applied modulation $\tau = 2 \pi / \Omega$. This assures that in both cases the equation (\ref{EqComp2}) for the covariance matrix has the same linear form, with $S$ being a periodic function of time (in the limit $t \gg 1/k$) and $N$ being a constant driving. The conclusions we derived in section \ref{SecRes1}, must therefore hold, at least qualitatively, also for the system studied in \cite{Mari1, Mari2}.

An interesting question now rises. What if the two modulations are applied together? Can they interfere, either constructively or destructively, and sensibly alter the one-modulation picture?

To get an answer, we consider a new composite system, described by the Hamiltonian
\begin{eqnarray}
	\hat{ \mathcal{H}} &= & \hbar \omega_C \hat a^\dagger \hat a + \frac{\hbar \omega_M}{2} \hat p^2 + \frac{\hbar \omega_M}{2}  \big(1 + \epsilon \cos (\Omega_1 t)\big) \hat q^2   \nonumber   \\
	&& - \hbar G_0 \hat a^\dagger \hat a  \hat q  \label{EqTwoHam} \\ \nonumber 
	&& + i \hbar E \big(1 + \eta \cos(\Omega_2 t + \phi) \big) (e^{- i \omega_L t} \hat a^\dagger - e^{i \omega_L t} \hat a). \label{hamtot}
\end{eqnarray}
Note that we explicitly introduced a relative phase $\phi$ between the two applied modulation: if we expect any interference, the properties of the system should indeed depend on this new variable.

The analysis presented in the previous sections is straightforwardly generalized to the present case, so we will skip directly to the results [details can be found however in the Appendix]. Taking the same parameters as in Sec.~\ref{SecRes1}, we choose the optimal modulation frequencies $\Omega_1=\Omega_2=2 \omega_M$ and fix $\epsilon=0.3$, $\eta=0.9$ (this is the same value used in \cite{Mari1}). These modulation strengths give comparable squeezing performances when considered singularly, and also assure that we are reasonably far from the instability region. To present the results, we plot the quantities introduced in Sec.~\ref{SecQuantumProp} against the relative phase $\phi$ in Fig. \ref{FigTwoModPhonons}.
An interference pattern is indeed evident and each of the above quantities oscillates between a minimum and a maximum as $\phi$ varies in the range $[0,2\pi]$. However, entanglement and quantum discord are affected very weakly and do not differ much from our initial one-modulation case. Besides we see that in order to generate quantum correlations, a modulation of the mechanical frequency is more suitable than a modulation of the driving laser amplitude. Adding the second modulation to the first is of little effect.

Squeezing generation instead, presents very interesting features. First, as we said, we choose two modulations that give comparable levels of squeezing when applied individually. Moreover, when applied together, they can strongly interfere. For example we see in Fig. \ref{FigTwoModPhonons}(d) that for a phase $\phi/\pi \sim 1.4$, $\Delta^2 q_{MIN}$ rises toward the threshold value $0.5$ and squeezing becomes weaker. Each modulation taken alone would generate more squeezing than the two combined: this is an unambiguous sign of a disadvantageous interplay. For a phase $\phi/\pi \sim 0.4$ we find instead a great advantage in applying two modulations: $\Delta^2 q_{MIN}$ is lowered to a value $\sim 0.18$, a considerable performance if compared to our initial one-modulation case where instabilities prevent us from reaching $\Delta^2 q_{MIN} < 0.17$. In fact, not only we attain the same high levels of squeezing, but we are also well inside the stability region, so that we could increase both $\epsilon$ and $\eta$ to perform even better. 

We also note that the optimal(worst) phase choice for squeezing generation also corresponds to maximum heating(cooling) of the mirror, as can be seen in Fig. \ref{FigTwoModPhonons}(a). This is another confirmation that parametric oscillation is indeed the main underlying mechanism: in fact not only does a stronger modulation enhance the generation of quantum effects, as inferred from equations (\ref{NEW}), it also pumps more energy into the system.

We then see how the interplay between two independent modulations can be carefully exploited to increase levels of squeezing in an optomechanical system.

\section{Conclusions}
\label{SecCon}

 We have studied in great detail the effect of periodic modulations on optomechanical systems and we have characterized several ways in which such modulations can be exploited to enhance relevant quantum properties including squeezing, entanglement and quantum discord.
While the idea that modulations can help accessing the quantum regime was already known from previous works \cite{Mari1, Mari2, Galve1}, we have proposed a new interpretation of this enhancement mechanism in terms of a resonance between the modulation frequency and the natural frequencies of the system. This simple model allowed us to prove the existence of an optimal modulation regime and to understand the arising of instability thresholds. Finally, we have analyzed the interplay of different modulations and we have found that constructive (destructive) interference effects may arise when they are applied simultaneously, causing a further enhancement (a suppression) of quantum effects. We believe that these results could lead further on toward the development of optimal control strategies.

\acknowledgments
We thank Rosario Fazio for useful discussions and comments. 
This work was supported by MIUR through FIRB-IDEAS Project No. RBID08B3FM.

\newpage

\appendix
\section{Asymptotic behaviors} \label{appa}

This section deals with  some technical aspects 
related with the asymptotic solutions of Eqs.~(\ref{EqSetMean}) and (\ref{EqComp2}), which define the quantum properties of the system. Here we discuss 
 the resonant mechanisms  which is responsible for  the enhancement of quantum effects at $\Omega\sim 2 \omega_M$, as well as the role of the relative phase in the interplay between different modulations. 
 We start in Sec.~\ref{single} by presenting a simple paradigmatic case which 
captures the main aspects of the resonance. 
Then, in section~\ref{SecAppGeneral}, we introduce the analytic framework which will be used to describe the dynamics of the system. Finally, sections~\ref{AppSingleMod}  and \ref{AppTwoMod} are devoted to analyze in details the asymptotic behavior of the system, in the one- and two-modulation scenario respectively.

\subsection{Single oscillator model}
\label{single}
As anticipated in the main text the evolution of the correlations matrix 
 describes the dynamics of  a
 multi-dimensional (classical) oscillator which 
evolves in presence of damping and external constant  driving (defined by the matrix $N$) 
 and which possesses characteristic frequencies (determined by $S(t)$)
  that are externally modulated at frequency $\Omega$ [these statements are explicitly verified in
 Secs.~\ref{SecAppGeneral},~\ref{AppSingleMod}  and \ref{AppTwoMod}]. 
To enlighten the role of the  modulation  in the evolution of the correlation functions it is hence worth focusing on 
 the simplest example of this sort. This is provided by a single parametric oscillator
whose  position $x$ evolves according to the equation
\begin{equation}
	\ddot x(t) = - \omega_0^2 [1 + \alpha \cos(\nu t)]\; x(t) - \gamma \dot x(t) + F\;,
	\label{EqParametricOscillator}
\end{equation}
with $\omega_0$ and $\nu$  being the characteristic and the modulation frequency, $\alpha$ being the amplitude of the modulation, $\gamma$ being the damping rate and $F$ the strength of a constant driving. 
For this simple scenario, two cases are possible. If $\alpha \gtrsim 2 \gamma / \omega_0$ (above-threshold condition), parametric modulation pumps energy into the system at a faster rate with respect to dissipation; the system increases its energy exponentially and is therefore unstable. If $\alpha \lesssim 2 \gamma / \omega_0$ (below-threshold condition), the system reaches a stationary regime, given by the balance of pumping and dissipation. We can then look for a stable solution of Eq.~(\ref{EqParametricOscillator}), assuming that $\alpha$ is small and treating the solution perturbatively, i.e. $x (t)= x^{(0)}(t) + \alpha x^{(1)}(t) + O(\alpha^2)$. To order zero in $\alpha$ the system is just a damped driven harmonic oscillator, which relaxes toward its equilibrium position $\bar{x}^{(0)} = \lim_{t\rightarrow \infty} x^{(0)}(t) = F/\omega_0^2$. To first order in $\alpha$, the long time solution is then given by
\begin{align}
	\ddot x^{(1)}(t) & = - \omega_0^2 x^{(1)}(t)  - \omega_0^2 \cos(\nu t)\;\bar{x}^{(0)}  - \gamma \dot x^{(1)}(t) .
	\label{EqParametricOscillatorPerturbative}
\end{align}
Therefore we see that the parametric modulation, for the below-threshold regime, can be mapped onto an effective external driving $F \cos(\nu t)$ and the solution is easily found to be
\begin{equation}
	x(t) \simeq  \frac{F}{\omega_0^2} + \alpha\;  f(\nu) \;  F \cos (\nu t + \phi),
	\label{EqParametricOscillatorPerturbative2}
\end{equation}
with $f(\nu) = 1/\sqrt{(\omega_0^2 - \nu^2)^2 + (\gamma \nu)^2}$ being the Lorentzian response function of a classical harmonic oscillator. Clearly the superimposed oscillation, which we remind is an effect of the parametric modulation, will be much greater near resonance with the natural frequency $\nu \sim \omega_0$ and for $\alpha$ just below the instability threshold. Going to second order in $\alpha$ yields small deviation from this picture and we can stop our qualitative analysis here. In summary, parametric modulation can controllably enhance oscillations of the system coordinates if two main conditions are satisfied: the modulation must not be too strong, otherwise the system becomes unstable, and an external (constant) driving must also be applied, otherwise the system relaxes to $x(t \rightarrow \infty) = 0$ (as from Eq.~\ref{EqParametricOscillatorPerturbative2} with $F=0$). We also stress out that, in the below-threshold regime, the resonance condition is given by $\nu \sim \omega_0$ (i.e. the modulation frequency should be the same as the natural frequency of the system) and not by $\nu \sim 2 \omega_0$, as is the usual case of exponential parametric amplification.

\subsection{General treatment of the modulated optomechanical system}
\label{SecAppGeneral}

 Turning back to Eq.~(\ref{EqComp2}), we will see that all conditions are indeed satisfied: the coefficient matrix $S(t)$ is periodically modulated over time, stability can be verified with a numeric solution and external driving is provided by the noise correlation function $N$. The above result implies that in the case of our multi-dimensional parametric oscillator, maximum enhancement of the oscillations is expected when
 $\Omega$ matches the characteristic frequencies that govern the dynamics of the correlation functions in absence of the modulation. The latter are defined by the matrix $S(t)$
 of Eq.~(\ref{EqComp2}) when $\epsilon =0$ (and $\eta=0$ in the two-modulation scenario). As mentioned in the text, at least for the figure of merit we are concerned about in the paper (i.e. $\Delta^2 q_{MIN}$, $\Delta^2 X_{MIN}$, $\mathcal{D}_{MAX}$, etc),
 the relevant frequency is indeed $\sim 2\omega_M$. 

We can thus generalize the simple model of Sec.~\ref{single} to the present case and reproduce the numerical results we found in the main text with a semi-analytic solution of Eqs. (\ref{EqSetMean}) and (\ref{EqComp2}), which we briefly sketch here. In doing so, we will also identify and comment the relevant points which are responsible for the behaviours observed in Figs. \ref{FigMaxPhonons} and \ref{FigTwoModPhonons}.

\subsubsection{Classical solution.}
\begin{figure}[t]
\includegraphics[width=0.3\textwidth]{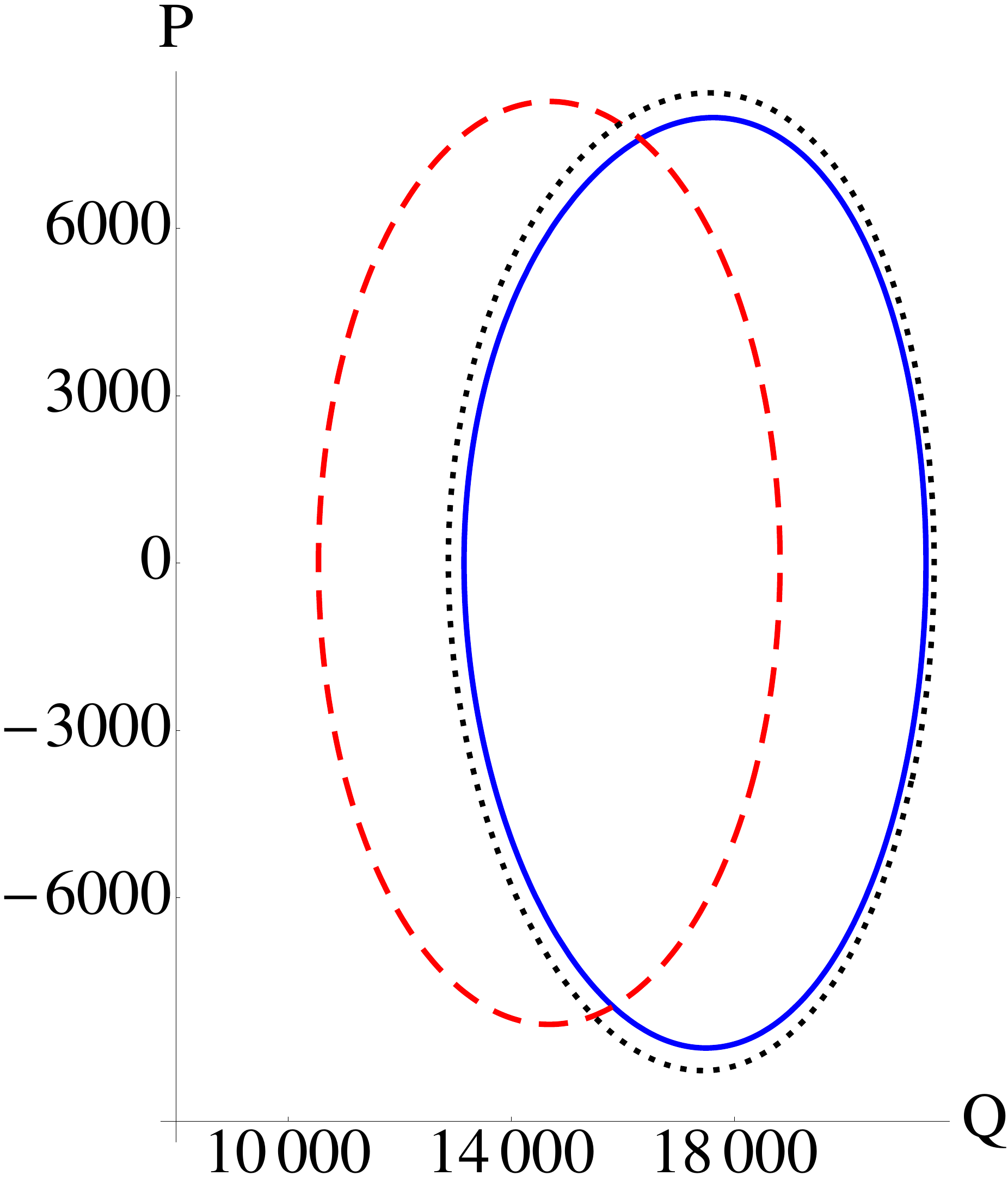}
\caption{First order (red dashed) and second order (blue solid) approximation to the classical position-momentum (Q-P) orbit of the mirror. The asymptotic numerical orbit (black dotted) is also plotted from $t=49 \tau$ to $t= 50 \tau$, with $\tau=2\pi/\Omega$ being the
period of the modulation. It is clear that a first order approximation is not enough and fails to describe the dynamics of the system. On the other hand, the second order approximation catches all relevant aspects and reproduces the correct behavior, at least on a qualitative level. Quantitative convergence to the numeric solution is found including higher order terms.}
\label{FigQPAnalytic}
\end{figure}

Let us start with Eq. (\ref{EqSetMean}) for the mean values which, for the sake of completeness, we report here for the general scenario defined by the Hamiltonian~(\ref{hamtot}) where both the frequency modulation~(\ref{EqDimHam}) and the amplitude modulation~(\ref{EqMariHam}) are activated, i.e. 
\begin{equation}
	\begin{cases}
	\partial_t  Q = \omega_M P,\\
	\partial_t  P = -\omega_M \;[1 + \epsilon \cos (\Omega t)]\; Q - \gamma_M P + G_0 |A|^2,\\
	\partial_t A = -(k + i \Delta_0) A + i G_0 A Q + E [1 + \eta \cos(\Omega t + \phi)],
	\end{cases}
	\label{EqSetMean111}
\end{equation}
having only  assumed  their frequencies  to be identical, i.e.  $\Omega_1=\Omega_2 = \Omega$. 
As anticipated in the text -- see Eq.~(\ref{ANapprox1}) --  we look for a perturbative solution in the modulations strengths $\epsilon$ and $\eta$, i.e\\
\begin{equation}
	\begin{cases}
	Q = Q^{(0)} + Q^{(1)} + Q^{(2)} + \dots,\\
	P = P^{(0)} + P^{(1)} + P^{(2)} + \dots,\\
	A = A^{(0)} + A^{(1)} + A^{(2)} + \dots,
	\end{cases}
	\label{EqClassicalExpansion}
\end{equation}\\
where, for instance, $Q^{(1)}$ is linear in $\epsilon$ and $\eta$, $Q^{(2)}$ is quadratic in $\epsilon$ and $\eta$ and so on (note that we can revert to the single modulation scenario simply by imposing $\eta=0$). At order zero we get\\
\begin{equation}
	\begin{cases}
	\partial_t \; Q^{(0)} = \omega_M \; P^{(0)},\\
	\partial_t \; P^{(0)} = -\omega_M \; Q^{(0)} - \gamma_M P^{(0)} + G_0 |A^{(0)}|^2,\\
	\partial_t \; A^{(0)} = -(k + i \Delta_0) A^{(0)} + i G_0 A^{(0)} Q^{(0)} + E.
	\end{cases}
	\label{EqClassicalZeroOrder}
\end{equation}\\
From the numeric simulation we know that this non-linear equation evolves toward a stable point ($\bar Q^{(0)}$, $\bar P^{(0)}$, $\bar A^{(0)}$) and by setting the derivatives to zero, we can find these asymptotic values. Next, at first order we get\\
\begin{widetext}
\begin{equation}
	\begin{cases}
	\partial_t \; Q^{(1)} = \omega_M P^{(1)},\\
	\partial_t \; P^{(1)} = -\omega_M \; Q^{(1)}  -\omega_M \; \epsilon \cos (\Omega t)\; \bar Q^{(0)} - \gamma_M P^{(1)} + G_0 \left( \bar A^{(0)} \right)^* A^{(1)} + G_0 \left( A^{(1)} \right)^* \bar A^{(0)},\\
	\partial_t \; A^{(1)} = -(k + i \Delta_0) A^{(1)} + i G_0 \bar A^{(0)} Q^{(1)} + i G_0 A^{(1)} \bar Q^{(0)} + E\; \eta \cos (\Omega t + \phi).
	\end{cases}
	\label{EqClassicalOrderOne}
\end{equation}\\
\end{widetext}
These are the equations of three coupled and forced harmonic oscillators, with forcing terms $-\omega_M \; \epsilon \cos (\Omega t)\; \bar Q^{(0)}$ and $E\; \eta \cos (\Omega t + \phi)$ that are purely oscillating. In addition, damping makes sure that the system is stable. The solutions are easily obtained in the form\\
\begin{equation}
	\begin{cases}
	Q^{(1)} (t) = q_1 e^{i \Omega t} +  q_1^* e^{-i \Omega t},\\
	P^{(1)} (t) = p_1 e^{i \Omega t} +  p_1^* e^{-i \Omega t},\\
	A^{(1)} (t) = a_1 e^{i \Omega t} +  a_2 e^{-i \Omega t},
	\end{cases}
	\label{EqClassicalOrderOneExpansion}
\end{equation}
with $q_1$, $p_1$, $a_1$, and $a_2$ being complex parameters which can be computed by replacing~(\ref{EqClassicalOrderOneExpansion}) into 
Eq.~(\ref{EqClassicalOrderOne}). 
Hence, to first order in $\epsilon$ and $\eta$, the effect of the modulation on classical values is to add an oscillating term with frequency $\Omega$ and mean value $0$. The amplitude of this oscillation clearly depends on the forcing term, i.e on the amplitudes $\epsilon$, $\eta$, on their relative phase $\phi$ and on the frequency $\Omega$ (via the oscillator response function). Finally, the equations for second order are\\
\begin{widetext}
\begin{equation}
	\begin{cases}
	\partial_t \; Q^{(2)} = \omega_M P^{(2)},\\
	\partial_t \; P^{(2)} = -\omega_M \; Q^{(2)}  -\omega_M \; \epsilon \cos (\Omega t)\; Q^{(1)} - \gamma_M P^{(2)} + G_0 A^{*(0)} A^{(2)} + G_0 A^{*(1)} A^{(1)} + G_0 A^{*(2)} A^{(0)},\\
	\partial_t \; A^{(2)} = -(k + i \Delta_0) A^{(2)} + i G_0 A^{(2)} Q^{(0)} + i G_0 A^{(1)} Q^{(1)} + i G_0 A^{(0)} Q^{(2)}.
	\end{cases}
	\label{EqClassicalSecondOrder}
\end{equation}\\
\end{widetext}
These are again the equations of three coupled and forced harmonic oscillators, but this time the forcing terms $-\omega_M \; \epsilon \cos (\Omega t)\; Q^{(1)}$, $G_0 A^{*(1)} A^{(1)}$ and $i G_0 A^{(1)} Q^{(1)}$ have also a constant part.  As for the first order corrections the solutions are easily obtained in the form\\
\begin{equation}
	\begin{cases}
	Q^{(2)} (t) = \bar{Q}^{(2)} + q_3 e^{i 2 \Omega t} +  q_3^* e^{-i 2 \Omega t},\\
	P^{(2)} (t) = \bar{P}^{(2)} + p_3 e^{i 2 \Omega t} +  p_3^* e^{-i 2 \Omega t},\\
	A^{(2)} (t) = \bar{A}^{(2)} + a_3 e^{i 2 \Omega t} +  a_4 e^{-i 2 \Omega t}.
	\end{cases}
	\label{EqClassicalSecondOrderExpansion}
\end{equation}\\
To second order in $\epsilon$ and $\eta$, the effect of the modulation on classical values is thus to add a constant shift and an additional oscillating term with frequency $2 \Omega$ and mean value~$0$. Higher orders can be processed in the same way but  for the parameter region we have selected in the main text, 
one can  limit the analysis to second order  since already at this point we get the correct result within a good degree of accuracy (see Fig \ref{FigQPAnalytic}). Full convergence of the approximation when higher orders are included can be seen from Fig \ref{FigMeanQP} in the main text, where we plot the numeric evolution of classical values $Q$ and $P$ and the analytic counterpart, computed up to order six.

\subsubsection{Linearized quantum solution.}
We now turn to Eq. (\ref{EqSetFluc}) for the quantum fluctuation, which we rewrite below
\begin{equation*}
	\partial_t {C} = S {C} + {C} S^\top + N.
\end{equation*}
Recall that the matrix $S(t)$ depends on $\epsilon$ and $\eta$ via the classical values and an additional explicit term $-\omega_M \; \epsilon \cos (\Omega t)$. If we make use of the approximate solution found before, we can thus identify a matrix $S^{(0)}$ independent of the perturbation, a matrix $S^{(1)}$ linear in $\epsilon$ and $\eta$ and a matrix $S^{(2)}$ quadratic in $\epsilon$ and $\eta$. Again, we look for a perturbative solution for the matrix $C$, i.e\\
\begin{equation}
	C = C^{(0)} + C^{(1)} + C^{(2)} + \dots.
	\label{EqFluctuationExpansion}
\end{equation}\\
where $C^{(1)}$ is linear in $\epsilon$ and $\eta$, $C^{(2)}$ is quadratic in $\epsilon$ and $\eta$ and so on. The calculations simply follow what we have done for the classical part. At order zero we get\\
\begin{equation}
	\partial_t \; C^{(0)} = S^{(0)} \cdot C^{(0)} + C^{(0)} \cdot S^{(0)\top} + N.
	\label{EqCorrelationsZeroCompact}
\end{equation}\\
This equation is linear and evolves toward a stable point $\bar C^{(0)}$, which we can find by setting the derivatives to zero. Next, at first order we get\\
\begin{equation}
	 \partial_t \; C^{(1)} = S^{(0)} \cdot C^{(1)} + C^{(1)} \cdot S^{(0)\top} + S^{(1)} \cdot \bar C^{(0)} + \bar C^{(0)} \cdot S^{(1)\top}.
	\label{EqCorrelationsOneCompact}
\end{equation}\\
These are the equations of sixteen coupled and forced harmonic oscillators, with forcing terms $S^{(1)} \cdot \bar C^{(0)} + \bar C^{(0)} \cdot S^{(1)\top}$ that are purely oscillating. Since $C$ is real, the solutions are easily obtained in the form\\
\begin{equation}
	C^{(1)}(t) = c_1 \; e^{i \Omega t} +  c_1^* \; e^{-i \Omega t}.
	\label{EqCorrelationsOrderOneExpansion}
\end{equation}\\
Hence, to first order in $\epsilon$ and $\eta$, the effect of the modulation on the correlations is to add an oscillating term with frequency $\Omega$ and mean value $0$. As in the case of an unidimensional resonator, the amplitude of this oscillation will be greater when the modulation frequency $\Omega$ is chosen in resonance with the eigenfrequencies of the normal modes. Finally, the equations for second order are\\
\begin{widetext}
\begin{equation}
	 \partial_t \; C^{(2)} = S^{(0)} \cdot C^{(2)} + C^{(2)} \cdot S^{(0)\top} + S^{(1)} \cdot C^{(1)} + C^{(1)} \cdot S^{(1)\top} + S^{(2)} \cdot \bar C^{(0)} + \bar C^{(0)} \cdot S^{(2)\top}.
	\label{EqCorrelationsSecondOrderCompact}
\end{equation}\\
\end{widetext}
These are again the equations of sixteen coupled and forced harmonic oscillators, but this time the forcing terms $S^{(1)} \cdot C^{(1)} + C^{(1)} \cdot S^{(1)\top}$ and $S^{(2)} \cdot \bar C^{(0)} + \bar C^{(0)} \cdot S^{(2)\top}$ have also a constant part. The solutions are easily obtained in the form
\begin{equation}
	C^{(2)}(t) = \bar C^{(2)} + c_{3} \; e^{i 2 \Omega t} +  c_{3}^* \; e^{-i 2 \Omega t}.
	\label{EqCorrelationsSecondOrderExpansion}
\end{equation}\\
Hence, as for the linear solutions, to second order in $\epsilon$ and $\eta$, the effect of the modulation on the correlations is to add a constant shift and an additional oscillating term with frequency $2 \Omega$ and mean value $0$.


\subsection{Asymptotic behavior in the single modulation regime} \label{AppSingleMod} 

\subsubsection{Classical solution.}
\label{AppSingleClassic}
We can come back to the single modulation scenario by putting $\eta = 0$ in the above analysis. By doing so, we can get approximate analytic expressions for the asymptotic mean values $A(t)$, $Q(t)$ and $P(t)$. However the complete formulas are too long to be reported here and we must limit ourselves to a semi-numeric expression, where we substitute all values as in Sec.~\ref{SecRes1} except for the interesting parameter $\epsilon$. For example we report the expression of the mirror position $Q(t)$\\
\begin{align}
	Q(t) & = 14684.7  -\epsilon^2 \; 2784.43 \nonumber \\
	& + \epsilon \Big( 4947.11 \; \cos(\Omega t) - 14.79 \; \sin(\Omega t) \Big) \nonumber \\
	& + \epsilon^2 \Big(164.97 \; \cos( 2 \Omega t) - 0.50\; \sin( 2 \Omega t) \Big).
	\label{EqQAnalytic1Mod}
\end{align}\\
As anticipated in the main text, to first order in $\epsilon$ the mean values have an asymptotic oscillatory behavior, which well describes the exact asymptotic solution. To be precise however, we cannot neglect the second order contributions: indeed, while second harmonic oscillations are one order of magnitude smaller, the constant shift is comparable to first order effects and must be taken in account.\\\\

\subsubsection{Linearized quantum solution.}
We can also look at the quantum properties of the system in the asymptotic regime, as a function of the modulation strength $\epsilon$. Fixing all other parameters to values in the text, we find for example the following expression for the number of phonons in the mirror $n_{phon}(t) \approx (C_{11}(t) + C_{22}(t) - 1)/2$\\
\begin{align}
	n_{phon}(t) & = 0.08  + \epsilon^2 \; 4.14 \nonumber \\
	& + \epsilon \Big( 0.14 \; \cos(\Omega t) - 0.01 \; \sin(\Omega t) \Big) \nonumber \\
	& - \epsilon^2 \Big(0.02 \; \cos( 2 \Omega t) - 0.21\; \sin( 2 \Omega t) \Big).
	\label{EqPhononsAnalytic1Mod}
\end{align}\\
For completeness we also report the expression for the single correlation $C_{11}$\\
\begin{align}
	C_{11}(t) & = 0.56  + \epsilon^2 \; 4.01 \nonumber \\
	& + \epsilon \Big( 0.28 \; \cos(\Omega t) - 1.63 \; \sin(\Omega t) \Big) \nonumber \\
	& + \epsilon^2 \Big(0.03 \; \cos( 2 \Omega t) - 0.20\; \sin( 2 \Omega t) \Big).
	\label{EqC11Analytic1Mod}
\end{align}\\
We see that $C_{11}(t)$ (and similarly $C_{22}(t)$) has strong oscillations in time proportional to $\epsilon$. Hence we can say, at least qualitatively, that squeezing will be dominated by first order effects in the range of values considered. On the contrary the number of phonons can be considered time-independent, with oscillations that are negligible if compared to the constant term. Moreover, we know that the mirror is cooled close to its ground state when the system is unmodulated. Therefore the number of phonons is strongly dependent on $\epsilon^2$, and the constant shift due to second order effects becomes quickly the dominant effect. These results agree very well with the numerical simulation summarized in Fig.~\ref{FigMaxPhonons}. 

We conclude this section with one last comment on why the number of phonons is constant in time. We know that the position $Q(t)$ and momentum $P(t)$ of the mirror oscillate with frequency $\sim \omega_M$ and a relative phase shift  of $\pi/2$ (slight modifications being induced by the interaction with the optical subsystem). In the same way $C_{11} = \left<\delta q^2 \right>$ and $C_{22} = \left< \delta p^2 \right>$ oscillate with twice this frequency, i.e. $\sim 2 \omega_m$, and with twice this relative phase shift , i.e. $\pi$. In turn, the two oscillations cancel each other out when summing $C_{11}$ and $C_{22}$, thus giving a time-independent number of phonons. In addition we see that the modulation is most effective on the mirror correlations when $\Omega \sim 2 \omega_m$, as stated before.


\subsection{Asymptotic behavior in the two modulation regime} \label{AppTwoMod}

\subsubsection{Classical solution.} 
We now reintroduce the second modulation and study the interplay between the two. Again we would like to fix all parameters except $\epsilon$, $\eta$ and $\phi$ to the values found in the main text. However, already at the classical level, expressions for  $A(t)$, $Q(t)$ and $P(t)$ tend to become rather long and complex since we have now 3 free parameters. Therefore we will substitute also the numerical values of $\epsilon$ and $\eta$ (values are found in Sec.~\ref{SecRes2}). This is not so bad: indeed recall that we are particularly interested in the dependence of quantum properties on the relative phase $\phi$. For example we report the expression of the mirror position $Q(t)$\\
\begin{align}
	Q(t) & = 17523.4 - 357.13 \; \cos(\phi) + 315.98 \; \sin(\phi) \nonumber \\
	& + 1484.13 \; \cos(\Omega t) - 4.43 \; \sin(\Omega t) \nonumber \\
	& + 2201.22 \; \cos(\Omega t + \phi) - 1870.83 \; \sin(\Omega t + \phi) \nonumber \\
	& + 14.84 \; \cos(2 \Omega t) - 0.04 \; \sin(2 \Omega t) \nonumber \\
	& + 22.62 \; \cos(2 \Omega t + \phi) - 18.78 \; \sin(2 \Omega t + \phi) \nonumber \\
	& + 113.53 \; \cos(2\Omega t + 2\phi) - 32.43 \; \sin(2 \Omega t + 2 \phi). 
	\label{EqQAnalyticPhase}
\end{align}\\
Without losing much time on the cumbersome formula above, we only point out that again first order effects are dominating (as we can see by comparing oscillations at $\Omega$ and oscillations at $2 \Omega$). However we also see that, already at the classical level, the the phase $\phi$ has a strong influence on the amplitude of oscillations. This is a clear sign that the phase plays indeed an important role in the system dynamics.\\\\

\subsubsection{Linearized quantum solution} 
\begin{figure}[ht]
\includegraphics[width=0.45\textwidth]{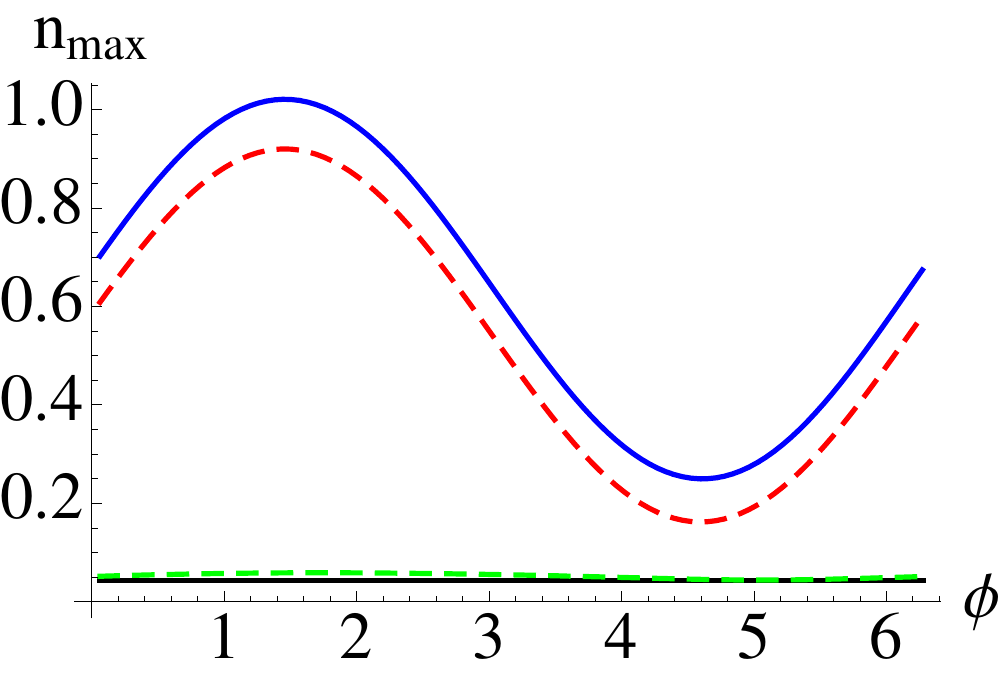}
\caption{In blue (upper solid) the maximum number of phonons, from Eq.~(\ref{EqPhononsPhase}), plotted against the relative phase $\phi$. Contributions due to different expansion orders are explicitly included: in black (lower solid) the number of phonons in the unmodulated case (from $\bar C^{(0)}$); in red (upper dashed) the second order time independent contribution (from $\bar C^{(2)}$); in green (lower dashed) the maximum over one period of the time dependent contributions (from first order $C^{(1)}$ and second order $C^{(2)}$). The total (blue) curve is equal to the sum of the other three curves.}
\label{FigPhononsAnalytic}
\end{figure}

We turn now to quantum properties of the system. Again we fix all parameters at the values used above, except for the relative phase $\phi$. To second order in the perturbation, we get expressions for the number of phonons and the correlation $C_{11}(t)$ that depends on the phase as\\
\begin{widetext}
\begin{align}
	n_{phon}(t)=\frac{1}{2} \Big( & 1.167 + 0.087 \; \cos(\phi) + 0.753 \; \sin(\phi) + 0.086 \; \cos(\Omega t) -0.005 \; sin (\Omega t) \nonumber \\
	& - 0.004 \; \cos(2 \Omega t) + 0.037 \; \sin(2 \Omega t) + 0.008 \; \cos(\Omega t + \phi) - 0.006 \; \sin(\Omega t + \phi) \nonumber \\
	& - 0.024 \; \cos(2 \Omega t + \phi) + 0.005 \; \sin(2 \Omega t + \phi) + O(10^{-5}) \Big),
	\label{EqPhononsPhase}
\end{align}
\begin{align}
	C_{11}(t)= & 1.09 + 0.02 \; \cos(\phi) + 0.39 \; \sin(\phi) + 0.08 \; \cos(\Omega t) -0.48 \; sin (\Omega t) \nonumber \\
	& + 0.002 \; \cos(2 \Omega t) - 0.02 \; \sin(2 \Omega t) + 0.39 \; \cos(\Omega t + \phi) - 0.05 \; \sin(\Omega t + \phi) \nonumber \\
	& + 0.001 \; \cos(2 \Omega t + \phi) - 0.002 \; \sin(2 \Omega t + \phi) + O(10^{-5}),
	\label{EqC11Phase}
\end{align}
\end{widetext}
where for brevity we have neglected the smallest terms.
Looking at expression~(\ref{EqPhononsPhase}) above, it is clear that the main effects of the modulation are contained in the first and third terms, other terms being an order of magnitude smaller than the two. The two bigger terms are both independent of time (similarly to the single modulation case), hence they must come either from the unmodulated solution $C^{(0)}$ or from the constant part of the second order solution $\bar C^{(2)}$. Again, since in the unmodulated scenario the mirror is very close to its ground state, we can reasonably assume that the matrix $C^{(0)}$ contributes in a negligible way. Therefore, the dependance of $n_{max}$ on the phase $\phi$ is almost entirely described by the matrix $\bar C^{(2)}$ and is a second order effect in the modulation strengths $\epsilon$ and $\eta$.
The maximum number of phonons (i.e. Eq.~(\ref{EqPhononsPhase}) maximized over one period $\tau=2 \pi/ \Omega$ of evolution), as well as the various contributions described here, are plotted in Fig. \ref{FigPhononsAnalytic}. We see that the analytic approximation correctly resembles the numeric solution (see Fig. \ref{FigTwoModPhonons}), the two differing only by a small constant shift which is due to higher order corrections. However, the qualitative behavior is fully understood already at the second order; therefore we do not report here explicitly higher orders contributions.

From Eq.~(\ref{EqC11Phase}), we see that $C_{11}$ (and similarly $C_{22}$) undergoes strong oscillations in time, reaching a minimum value that depends much on the phase $\phi$. This tells us that mechanical squeezing will have a very similar behavior and hence will also depend strongly on $\phi$.

Entanglement and quantum discord have instead a much smaller response. Indeed, both quantities are computed using all entries of the matrix $C$ (and not only two entries as in the case of phonons). Each entry will have an expression similar to~(\ref{EqC11Phase}) and made of three parts: a constant part, a time independent part which oscillates with the relative phase $\phi$ and a time dependent part. This can be written in the general form\\
\begin{equation}
	C_{ij} = A_{ij} + B_{ij} \; \cos(\phi + \varphi_{ij}) + C_{ij}(t)
	\label{EqPrototypeExpressionCij}
\end{equation}\\
with $A_{ij}$, $B_{ij}$ and $ \varphi_{ij}$ constants. In general, the oscillations $B_{ij} \; \cos(\phi + \varphi_{ij})$ will be out of phase with one another. Also the time dependent parts $C_{ij}(t)$ will be generally oscillating out of phase. Therefore, summing many entries together, these parts will cancel each other out (as a sum of incoherent waves) and only the constants $A_{ij}$ survive to play a relevant role. From this handwaving reason, we expect that entanglement and quantum discord should be quite insensitive to phase $\phi$, as is the case in Fig. \ref{FigTwoModPhonons}.\\

\begin{thebibliography}{99}
\bibitem{NOTAEXP}
 Realistic dissipation and noise processes for the levitated sphere models of Refs.~\cite{Chang1,Romero1,Li1} are somewhat different from the general description presented here: indeed the sphere is mechanically isolated from any thermal reservoir, so that thermalization by contact is absent; an important disturbance effect comes instead from scattering of photons out of the cavity \cite{Romero1}, which is not included in our equations. Hence, realistic and quantitative predictions for the systems of Refs.~\cite{Chang1,Romero1,Li1} using our calculations
should be made only after 
a careful revisitation of the model. However the generation of quantum effects is entirely due to the optomechanical interaction, while noise and dissipation just impose a boundary on how good the performance can be. In the low noise limit, the results should qualitatively remain the same.
\bibitem{NOTE111} Notice that an analogous phase-locking mechanism occurs also for the
evolution of the mean variables $Q(t)$, $P(t)$ and $A(t)$. Indeed as evident from Eq.~(\ref{EqSetMean})
the oscillator $Q(t)$, $P(t)$ of natural frequency $\omega_M$ evolves under a parametric modulation of the phase of frequency $\Omega$. The parametric enhancement in this case however has not direct
consequences on the evolutions of the correlation matrix ${C}$, but for the fact that the intensities of the
modulations of the latter tend to be increased.
\end{thebibliography}
\end{document}